\documentclass[12pt]{article}
\usepackage{amssymb,amsmath}
\usepackage{bm} 
\usepackage{booktabs} 
\usepackage{array}
\usepackage{latexsym}
\usepackage{graphicx}
\usepackage{color}
\usepackage{slashed}
\usepackage{datetime}
\usepackage[nosort]{cite}
\usepackage{verbatim}
\usepackage{enumerate}
\usepackage{chngpage} 
\allowdisplaybreaks
\usepackage{psfrag}
\usepackage{mathtools}
\usepackage{mciteplus}
\usepackage{dsfont}
\usepackage{multirow} 
\usepackage{bbm} 

\usepackage[colorlinks=true,      linkcolor=blue,      urlcolor=blue,      
            filecolor=blue,      citecolor=blue,       pdfstartview=FitH,     
						pdfpagemode=UseNone,      bookmarksopen=true]{hyperref}  
\usepackage[all]{hypcap}     

\usepackage{tikz}
\usepackage{pgfplots}
\usepackage{amsmath}
\pgfplotsset{compat=1.16}
\usepackage{amsmath}

\newcommand{\bra}[1]{\langle#1|}
\newcommand{\ket}[1]{|#1 \rangle}

\def\({\left(}
\def\){\right)}
\def\[{\left[}
\def\]{\right]}

\def\pd{{\partial}}

\def\barray{\begin{array}}
\def\earray{\end{array}}
\def\be{\begin{equation}}
\def\ee{\end{equation}}
\def\bea{\begin{eqnarray}}
\def\eea{\end{eqnarray}}
\def\bal{\begin{align}}
\def\eal{\end{align}}
\def\nn{\nonumber}

\numberwithin{equation}{section} 
\allowdisplaybreaks  

\makeatletter
\g@addto@macro\bfseries{\boldmath}
\makeatother

\definecolor{cardinal}{rgb}{0.6,0,0}
\definecolor{darkgreen}{rgb}{0,0.4,0}
\definecolor{golden}{rgb}{0.92, 0.7, 0}
\definecolor{midnight}{rgb}{0, 0, 0.5}
\definecolor{darkblue}{rgb}{0, 0, 0.7}
\definecolor{purple}{rgb}{0.5, 0, 0.5}

\def\tr{\mathop{\mathrm{Tr}}\nolimits}
\def\str{\mathop{\mathrm{STr}}\nolimits}

\def\ket#1{|{#1}\rangle}

\def\twopia{2\pi\alpha'}

\usepackage{epsf}
\usepackage{cite}
\usepackage{fancyhdr}
\usepackage{bm}
\usepackage{enumerate} 

\usepackage{empheq} \empheqset{box=\fbox}

\numberwithin{equation}{section}  
\allowdisplaybreaks

\begin{document}


\begin{flushright}
%

\end{flushright}

\vspace{3mm}

\vspace{3mm}

\begin{center}

{\huge {\bf The local supersymmetries\\ of a stack of branes}}

\vspace{14mm}

{\large
\textsc{Rapha\"el Dulac $^a$, Yixuan Li $^{b,c}$}}
\vspace{12mm}

\textit{$^a$	 Institut de Physique Th\'eorique, \\  
Universit\'e Paris-Saclay, CNRS, CEA,\\
	Orme des Merisiers, Gif-sur-Yvette, 91191 CEDEX, France  \\}  
\vspace{6mm}

\textit{$^b$	Dipartimento di Fisica e Astronomia “Galileo Galilei”, Università di Padova,\\ Via Marzolo 8, 35131 Padova, Italy \\} 
\vspace{0.5em}
\textit{$^c$	INFN, Sezione di Padova,\\ Via Marzolo 8, 35131 Padova, Italy\\}  
\medskip

\vspace{4mm} 
%

{\footnotesize\upshape\ttfamily  raphael.dulac @ ipht.fr , yixuan.li @ pd.infn.it} \\
\vspace{13mm}
 

\end{center}

\begin{adjustwidth}{10mm}{10mm} 

\begin{abstract}
\vspace{1mm}
\noindent

Local supersymmetries have been a guiding principle to construct smooth horizonless supergravity solutions. By computing brane densities from the brane low-energy effective action, we classify the BPS solutions that have the maximal local supersymmetry structure, with sixteen supersymmetries. We find that whenever the gauge group is broken to its maximal abelian subgroup, the brane construction becomes purely geometric and one can identify the preserved local supercharges. This class of solution includes two- and three-charge monopoles, and polarised branes. On the contrary, we show that instantons involve pure non-abelian degrees of freedom that cannot have a geometric interpretation. Hence, we conjecture that such instantons cannot be fully captured by supergravity.

\end{abstract}
\end{adjustwidth}

\thispagestyle{empty}
\clearpage



\baselineskip=14.5pt
\parskip=3pt

\tableofcontents

\baselineskip=15pt
\parskip=3pt

\clearpage

\section{Introduction}
In string theory, the microscopic degrees of freedom of supersymmetric black holes can be identified with bound states of branes and strings, and counted reliably in the limit where both the string coupling and gravitational interactions vanish\cite{Strominger:1996sh,Sen:1995in}. However, a fundamental question remains: what is the fate of these microstates once the coupling is finite and the classical black-hole solution emerges? In this regime, gravity becomes strong enough to support horizons and singularities, yet the microscopic structure should persist in some form.

There are essentially two possibilities. One possibility is that, at the scale of the horizon, all microstates become effectively indistinguishable, differing only by corrections that are exponentially suppressed in the entropy. In this scenario, one would need to go beyond the supergravity approximation and use the full  string theory to resolve individual microstates. The other, as argued by \cite{Mathur:2002ie,Mathur:2009hf,Almheiri:2013hfa,Bena:2022ldq,Mathur:2024ify}, is that the microstate solutions differ to the classical black-hole solution at the scale of the horizon. There seems to be evidence from string theory for the second possibility: there exists horizonless solutions with same mass, charge and angular momentum as the black hole, and such horizonless solutions have been found in many classes of black-hole solutions \cite{Lunin:2001fv,Lin:2004nb,Bena:2005va,Bena:2016ypk,Ganchev:2021pgs,Bah:2023ows} (see \cite{Bena:2025pcy} for a recent review).  However, the interpretation of the horizonless solutions that have been constructed is unclear. First, there is the possibility that such solutions are atypical black-hole microstates, and that the vast majority of black-hole microstates are indistinguishable from the black-hole solution at the horizon scale. Second, it might be that these solutions are not even part of the ensemble of states that constitute the black hole, and instead describe collectively a thermal ensemble of gravitons or of an exotic composite object \cite{Choi:2025lck,Larsen:2025jqo}. 

In order to solve the puzzle, one avenue is to track the states from the weak-coupling regime describing the brane physics to the strong-coupling regime where supergravity is well-defined.
Following the work of Strominger and Vafa \cite{Strominger:1996sh} which counts the entropy of black holes in the D1-D5-P frame, the standard lore is to track the states corresponding to propagating waves on the moduli space of instantons. The instantons describe the physics of D1 branes inside D5 branes with a common spatial direction, and the waves propagate in this common direction. More recently, a parallel effort has emerged in the D2-D4-P duality frame, where momentum-carrying states are instead constructed as waves on the moduli space of monopoles \cite{Bena:2022wpl}. Here, the monopoles describe the geometric structures taken by orthogonal D2 and D4 branes, with a common spatial direction. As in the D1-D5-P system, the D2 and D4 branes extend only in the compact spatial dimensions. As such, waves on D2-D4 monopoles should in principle have the same chance to describe black-hole microstates as those on D1-D5 instantons.

Tracking each microstate up to the supergravity regime, even for an extremal and supersymmetric black hole, is technically difficult, and might be fundamentally impossible. But perhaps the best chance for this endeavour to succeed is to make use of a special property of branes. When branes of different types combine together in order to form 1/4- and 1/8-BPS bound states, the bound state is locally 1/2-BPS: one can zoom on every location of the bound state, and it will look like a flat brane with sixteen supersymmetries \cite{Bena:2022wpl,Bena:2011uw,Bena:2022fzf}. For instance, a D1 brane ending on a D3 brane pulls the worldvolume of the D3 brane, and the resulting geometry is a ``BIon spike'' \cite{Callan:1997kz} which has sixteen local supersymmetries, even if the D1-D3 system preserves only eight supersymmetries globally. This `local supersymmetry enhancement' is actually at the heart of the techniques to generate horizonless solutions in the supergravity regime \cite{Bena:2022sge}.

This special property has been the focus of some recent works. The local supersymmetry enhancement structure of 1/4-BPS and 1/8-BPS systems has been classified for off-shell solutions in \cite{Eckardt:2023nmn}. And it was shown that all the solutions of the abelian Born-Infeld action have this property \cite{Bena:2024oeq}. In this article we propose to study the supersymmetry enhancement in the much richer class of the non-abelian generalisation of Born-Infeld action. A priori, whether one finds sixteen local supersymmetries or not depends on the duality frame and the types of branes that are involved (like D1-D5-P or D2-D4-P). We will show that local supersymmetry happens on shell in most of the duality frames.

Ultimately, our goal is to test whether the geometric structures emerging from brane interactions at weak string coupling can survive into the regime of finite gravitational coupling where the classical black-hole solution exists, and give rise to smooth, horizonless microstate geometries. In this work, we focus on $\alpha'$ corrections but with gravity turned off. We study the first $\alpha'$ correction of the unknown non-abelian Born-Infeld plus Wess-Zumino action. Since we study supersymmetric solutions, we expect our computations to remain valid to all orders in $\alpha'$. Here, at weak coupling, the geometric structures are encoded in the worldvolume theory of the branes. We will show many examples where the branes exhibit local supersymmetry enhancement. One of them is the D2-D4-P system with waves on D2-D4 monopoles that we described earlier. The waves are carried by excitations that are purely in the compact spatial dimensions.

By contrast, we show that the brane densities arising from D0-D4 (\textit{resp.} D1-D5) instantons with dipolar D2 (\textit{resp.} D3) branes do not preserve sixteen local supersymmetries. Therefore, this suggests that finding horizonless microstates that are the backreaction of momentum waves on the D1-D5 instanton moduli space may not be possible in the D1-D5 frame.\footnote{It is possible to polarize the D1-D5 branes into a Kaluza-Klein monopole with one spatial extension in the non-compact spatial dimension \cite{Bena:2011uw}, add momentum on top, and construct a supergravity solutions with the corresponding dipole charges \cite{Bena:2016ypk}. However, we are interested in momentum-wave excitations in the compact dimensions, as these are expected to lead to more typical states \cite{Eckardt:2023nmn,Bena:2022wpl,Bena:2022sge}.} Contrasting with the sixteen local supersymmetries found in the monopoles of the D2-D4 frame, our work suggest that the ability to find horizonless microstates with microstructure in the internal dimensions depends on the duality frame.

The paper is organised as follows. In Section \ref{sec:brane density formalism} we review the methods used to probe brane densities, employing both the `Taylor-Van Raamsdonk' procedure and $\kappa$-symmetry. Then we demonstrate in Section \ref{sec:monopoles} that the most general two-charge monopoles of $\mathrm{U}(N)$ and $\mathrm{SU}(N)$ Yang-Mills all have sixteen local supersymmetries. This result is extended in Section \ref{sec:brane_polarization}, where we establish the same for branes that polarize into higher-dimensional ones. In Section \ref{sec:instantons}, we argue that instantons do not have this structure. In Section \ref{sec:3-charge_monopole} we examine the case where a momentum charge is added to monopoles, and prove that the solution preserves the sixteen-local-supercharge structure. This supports the idea that monopole-based constructions (like the supermaze \cite{Bena:2022wpl}) inherit this structure, suggesting they may be constructible within supergravity. Finally, in Section \ref{sec:conclusion}, we conclude with speculative remarks on the duality-frame dependence of the ability to find typical horizonless solutions.

\section{How to compute brane densities?}
\label{sec:brane density formalism}

The low-energy effective action of a single probe D$p$-brane in the background of a type-II supergravity solution is described by the Born-Infeld action, accompanied by a topological Wess-Zumino term\cite{Simon:2011rw}. The various gauge fields on the worldvolume of the D$p$-brane couple to the bulk supergravity fields via the Wess-Zumino action, revealing the possibility for a D$p$-brane to carry lower-dimensional branes. 

For a stack of $N$ D$p$-brane, the gauge group is enhanced from $\mathrm{U}(1)^N$ to $\mathrm{U}(N)$ because modes corresponding to open strings stretching between different branes become massless when all the branes lie at the same position. The precise definition of the non-abelian version of the Born-Infeld action is poorly understood, mainly because of the difficulty to differentiate between higher-derivative corrections and $\alpha'$ corrections. Indeed, both types of corrections are mixed up through the identity $[D_i,D_j]F_{kl}=-\frac{i}{2}[F_{ij},F_{kl}]$ \cite{Tseytlin:1997csa} (the field strength is proportional to $\alpha'$). While the Born-Infeld action accounts for all $\alpha'$ corrections, only the first-order approximation of the non-abelian Born-Infeld action is well-understood and described by maximally supersymmetric super Yang-Mills theory. Following work of Seiberg \cite{Seiberg:1997ad} and Sen \cite{Sen:1997we}, Taylor and Van Raamsdonk \cite{Taylor:1999gq,Taylor:1999pr} have shown how to identify the brane charges (up to their linear dependence) from the Wess-Zumino-like action. The task is more complicated than for the abelian case, since from a stack of D$p$-brane, higher-dimensional branes can emerge thanks to the coupling with non commuting scalar fields. Consequently, the stack of branes can couple to both higher- and lower-dimensional branes. Before explaining the Taylor-Van Raamsdonk procedure, let us first briefly review the abelian case.

\subsection{Brane charges for a single brane}
\label{ssec:brane charges single stack}

The bosonic part of the brane low-energy effective action (only massless degrees of freedom) of a single probe brane is divided into two pieces: the Born-Infeld term and the Wess-Zumino term:
\be \label{probe_action}
S_\mathrm{probe}= S_\mathrm{DBI} + S_\mathrm{WZ} \,,
\ee
with
\begin{align}
S_\mathrm{DBI}= & -T_{\mathrm{D}p}\int \mathrm{d}^{p+1}\sigma \, e^{-\phi} \sqrt{-\mathrm{det}\left(g_{ij}+ (2\pi\alpha')F_{ij}-B_{ij}\right)}  \\
S_\mathrm{WZ} =& -T_{\mathrm{D}p} \int \mathrm{d}^{p+1}\sigma \, e^{(2\pi\alpha')F-  B}\wedge \( \sum_{l}C^{(l)} \) \,. 
\end{align}

In these expressions, $g_{ij}\,, B_{ij}$ are the pull-backs of the metric and of the NS-NS field on the world-volume of the brane. $F$ is the field strength of the $\mathrm{U}(1)$ gauge field, and  $(2\pi\alpha')F-B$ is the gauge-invariant quantity. The tension of the D$p$-brane is $T_{\mathrm{D}p}=\frac{1}{(2\pi)^p g_s l_s^{p+1}}$. Throughout the paper, we will call $S_\mathrm{probe}$ the ``Born-Infeld action'' through misuse of language.

The Wess-Zumino term should be understood as integrating over all $(p+1)$-forms arising from the expansion of the exponential. For simplicity, we focus on geometries without the NS-NS field, $B_{\mu\nu}$, and adopt the static gauge in the following discussion. Perhaps the cleanest approach for identifying the brane charges on the worldvolume of the brane is to use the total action, including the fermionic degrees of freedom, and extract the charges via $\kappa$-symmetry \cite{Bena:2024oeq}. This method directly connects the fields to a traceless involution (the $\kappa$-symmetry projection involution) from which the brane densities can be determined. However in this subsection, we simply read off the charges directly from the Wess-Zumino action. To build intuition, we provide a brief dictionary for reading off the charges from the Wess-Zumino action.

We consider the low-energy action of a D$p$-brane, $p\ge2$, and turn on one `magnetic' field $F_{12}$, which couples to the $C_{03\dots p}$ through the Wess-Zumino action:
\begin{equation}
    S_\mathrm{WZ} \supset -T_{\mathrm{D}p} \int \mathrm{d}^{p+1}\sigma F_{12}\wedge C_{03\dots p}\,.
\end{equation}
Thus, turning on a magnetic field $F_{12}$ on the worldvolume of the D$p$-brane is equivalent to having a D$(p-2)$-brane living on the brane on all the worldvolume directions orthogonal to $\{1, \,2\}$, and the brane density is given by $F_{12}$, see \textit{e.g.} \cite{Simon:2011rw}. The story generalises for $F_{ij}$, where $i,j$ are any \textit{spatial} worldvolume indices. Similarly, turning on two constant magnetic fields $F_{12}$ and $F_{34}$ is equivalent to having orthogonal D$(p-2)$ branes and D$(p-4)$ branes dissolved in the D$p$-brane:
\begin{equation}
    S_\mathrm{WZ} \supset -T_{\mathrm{D}p} \int \mathrm{d}^{p+1}\sigma \, \( F_{12}\wedge C_{03\dots p}\,+F_{34}\wedge C_{012\, 5\dots p}+F_{12}\wedge F_{34}\wedge C_{05\dots p}\) \,.
\end{equation}

In a similar fashion, we can turn on an `electric' field $F_{0i}$ which indicates the presence of fundamental strings along $i$. However, the density of the fundamental strings is not directly $F_{0i}$, but rather its conjugate variable, $\frac{\partial \mathcal{L}}{\partial F_{0i}}$ \cite{Simon:2011rw,Bena:2024oeq}. To linear order in the field strength (and in $\alpha'$), $\frac{\partial \mathcal{L}}{\partial F_{0i}}=F_{0i}+\mathcal{O}(F^2)$. The Taylor-Van Raamsdonk procedure will capture only the leading contribution in the density of fundamental strings, $F_{0i}$.

The simultaneous presence of an `electric' field $F_{0i}(\sigma)$ and a `magnetic' field $F_{ik}(\sigma)$ induces a local momentum current along the $k^\mathrm{th}$ direction, $F_{0i}F_{ik}$. This is the analogous of the Poynting vector $\vec{E}\wedge\vec{B}$ in electromagnetism.

When a scalar field is involved, it indicates that the brane has legs in an other transverse direction. For concreteness, take a scalar field $\Phi$ (which represents a transverse direction that we label as $z$), the coupling to the Wess-Zumino action is given by:
\begin{equation}
    S_{\mathrm{WZ}} \supset \int \mathrm{d}^{p+1}\sigma \, \epsilon^{\alpha_1 \dots\alpha_p \mu} \, C_{\alpha_1 \dots\alpha_p z } \, \partial_{\mu} \Phi \,.
\end{equation}
This is simply the pull-back of the R-R form on the brane. This emphasizes that turning on a scalar on the Born-Infeld action implies the presence of a brane density in the transverse direction with density $\partial_{\mu}\Phi$. If $\mu=0$, $\partial_\mu\Phi$ is a velocity, which couples to a momentum density with charge $\frac{\partial \mathcal{L}}{\partial(\partial_0\Phi) }\,$.

\subsection{Brane charge for a stack of multiple branes}
\label{ssec:TVR_formalism}

The low-energy effective action \eqref{probe_action} describing a stack of $N$ coincident branes is given, to leading order in $\alpha'$, by maximally supersymmetric $\mathrm{U}(N)$ Yang-Mills theory. In the abelian case, the Wess-Zumino coupling allows a single brane to interact only with lower-dimensional branes. However, in its non-abelian version, the transverse scalar fields become $\mathrm{U}(N)$-valued, and their non-commutativity allows couplings to higher-dimensional branes. This interaction can lead to lower-dimensional branes polarizing into higher-dimensional ones, as discovered by Myers \cite{Myers:1999ps}. In what follows, we focus on supersymmetric configurations of the Yang-Mills action, which are expected to persist as exact solutions of the full non-abelian Born-Infeld theory, much like in the abelian case.

In order to determine the brane densities, we must understand how the fields in the Yang-Mills action couple to the bulk supergravity fields. Such couplings have been obtained at linear order by Taylor and Van Raamsdonk \cite{Taylor:1999gq,Taylor:1999pr}. For BPS solutions, the linear order should be sufficient to measure completely the brane densities. First, we define the field $\mathcal{C} \equiv \exp(-B)\wedge\sum_l C^{(l)}$, where the $C^{(l)}$ refer to the R-R fields, while $B$ is the NS-NS two-form. The non-abelian version of the Wess-Zumino action is given by \cite{Taylor:1999pr}:
\begin{equation}
    S^{\mathrm{D}p}_\mathrm{WZ} = T_{\mathrm{D}p} \int \mathrm{d}^{p+1}\sigma \epsilon^{a_0 \cdots a_p} \sum_{q} \sum_{n=\max(0,q-p-1)}^{\min(q,9-p)} c^p_{q,n} \operatorname{STr} \left\{ {\cal C}^{(q)}_{ a_0 \cdots a_{q-n-1} i_1 \cdots i_n} F^{\,n+(p-q+1)/2}_{( a_{q-n} \cdots a_p i_1 \cdots i_n )} \right\} \,.
\label{Taylor-VRaamsdonk}
\end{equation}
The indices $a_k$ are the indices of the D$p$-brane's worldvolume (including time), while the indices $i_k$ denote its transverse coordinates. Wedge products are implicit.
Essentially, \eqref{Taylor-VRaamsdonk} tells that RR and NS-NS fields, combined into ${\cal C}^{(q)}_{ a_0 \cdots a_{q-n-1} i_1 \cdots i_n}$, couple to an expression involving the ``generalised'' field strength, $F$. The field strength is ``\textit{generalised}'' because its indices are defined to be ten-dimensional instead of being only on the brane worldvolume: Along the worldvolume indices, $F_{ab}$ is the usual field strength on the brane worldvolume; along mixed indices, $F_{ai}$ is $F_{ai}\equiv D_aX^i$;  with transverse indices, $F_{ij}$ is defined as $F_{ij}\equiv i [X^i,X^j]$.
The number $n+(p-q+1)/2$ in the `exponent' of $F$ denotes the number of times $F$ appears in that expression. Since the elements involving the $F$s do not commute, one performs a symmetrised trace, $\operatorname{Str}\{\cdot\}$, over the $F^{\,n+(p-q+1)/2}$.
The numerical coefficients $c^p_{q,n}$ are given by $c^p_{q,n} = \frac{(-1)^{n(n-1)/2} (p + 2n - q)!! }{ n! \, (q-n)! \, (n+p-q+1)!}$.

For example, the RR fields $C^{(q+1)}_{ a_0 \cdots a_{q-n} i_1 \cdots i_n}$ couple to an expression, $J_{\mathrm{D}q}^{ a_0 \cdots a_{q-n-1} i_1 \cdots i_n}$, which we take as that of the brane density of D$q$ branes along the directions $(a_0, \dots, a_{q-n}, i_1, \dots, i_n)$.

Nevertheless, we can and also, and will use an equivalent prescription to read off the brane densities. The Wess-Zumino action for a stack of D$p$-branes is given by \cite{Myers:1999ps}:
\begin{equation} \label{non-commutative_WZ}
    S_{\mathrm{WZ}}= T_{\mathrm{D}p}\int \mathrm{d}^{p+1}\sigma  \str\left(e^{i 2\pi\alpha' i_{\Phi}i_{\Phi}}\left(\sum_n C^{(n)}e^{-B}\right)e^{2\pi\alpha' F}\right)\,,
\end{equation}
where $i_{\Phi}i_{\Phi}\, C^{(n)}$ is defined to be:
\begin{equation}
    i_{\Phi}i_{\Phi}\, C^{(n)}=\frac{1}{2}[\Phi_i,\Phi_j]\, C^{(n)}_{0\dots (n-3)\,ij}\,.
\end{equation}
For example, in \eqref{non-commutative_WZ}, through the term
\begin{equation}
    i_{\Phi}i_{\Phi}\, C^{(p+3)}=\frac{1}{2}[\Phi_i,\Phi_j]\, C^{(p+3)}_{0\dots p\,ij}\,
\end{equation}
the field $C^{(p+3)}$ appears in the worldvolume action of the D$p$-brane. The symmetry between $e^{i 2\pi\alpha' i_{\Phi}i_{\Phi}}$ and $e^{2\pi\alpha' F}$ in the expression \eqref{non-commutative_WZ} highlights that lower-dimensional branes can source higher-dimensional ones through non-commuting scalar fields, $\{\Phi_i\}$, in the same way higher-dimensional branes can source lower-dimensional brane charges through $F$. The $e^{i 2\pi\alpha' i_{\Phi}i_{\Phi}}$ operator is always trivial in the abelian Born-Infeld action.

\subsection{Local supersymmetries, main branes and glues}

The purpose of this section is to review the framework for understanding local supersymmetry structures; we follow the approach outlined in \cite{Bena:2022wpl,Li:2023jxb,Bena:2024oeq}.

Type-II string theory vacua preserve 32 supersymmetries. Introducing additional branes, fundamental strings, or Kaluza-Klein momentum modes (which we will collectively refer to as “branes”) breaks a part of these supersymmetries. The preserved supersymmetries are determined by the BPS conditions, which manifest as solutions to the Killing-spinor equations. To each type of brane is associated a traceless involution, $P$, that characterises its supersymmetry properties. The BPS equations are a projection condition for the supersymmetry parameter $\varepsilon$:
\begin{equation}
    P\varepsilon=\varepsilon\,, \quad \text{or equivalently,} \quad \frac{1}{2}(1-P) \varepsilon =0 \,.
\end{equation}
In other words, the supersymmetries which are preserved in the presence of the brane belong to the kernel of the projector $\Pi\equiv \frac12(1+P)$. The complete list of brane involutions can be found in the Appendix \ref{appendix:projectors}.

Several types of branes are said to have ``compatible'' supersymmetries when their characterising involutions, $\{P_i\}$, commute. Indeed, in such cases, one can co-diagonalise the projectors, and the supersymmetries $\varepsilon$ that are preserved by that collection of branes lie in the intersection of the kernels, $\cap_i\ker(\Pi_i)$. The dimension of $\cap_i\ker(\Pi_i)$ counts the number of preserved supersymmetries.

But it is also possible to bound branes that seem non-compatible. Bounding a set of branes characterised by their involutions, $\{P_j\}$ lead to the projection condition:
\begin{equation}
    \Pi\,\varepsilon \equiv \frac{1}{2}(1+\alpha_1P_1+...+\alpha_nP_n)\,\varepsilon=0\,,
\end{equation}
where the coefficients $\alpha_i=\frac{j_i}{\mathcal{H}}$ are the ratios between the charge density of the brane type corresponding to $P_i$ and the energy density of the branes.

For generic values of the coefficients $\alpha_i$, $\Pi$ is not a projector (\textit{i.e.} $\Pi^2=\Pi$). But when $\Pi$ is a projector, then the dimension of its kernel is automatically sixteen (since $\tr P_i=0$ and $\tr \Pi = 16$), and so there are sixteen preserved supersymmetries. Physically, having sixteen supersymmetries for a system of branes is a rather special property. For example, D0, D2, D4 and D6 branes can extend on various cycles of a $\mathbb{T}^2\times\mathbb{T}^2\times\mathbb{T}^2$; the quantity of the branes can be chosen precisely such that the mirror dual is a single stack of D3 branes on $\mathbb{T}^2\times\mathbb{T}^2\times\mathbb{T}^2$, for which case sixteen supersymmetries are preserved \cite{Calderon-Infante:2025pls}.

In general, the set of $\{\alpha_i\}$ for which $\Pi$ is a projector is not unique and can be parametrised by continuous parameters. As such, the coefficients $\{\alpha_i\}$ can be functions in terms of spatial or worldvolume coordinates, $\vec{x}$:
\begin{equation} \label{projector_local}
    \Pi(\vec{x}) \, \varepsilon(\vec{x})=\frac{1}{2}\left(1+\alpha_1(\vec{x})P_1+...+\alpha_n(\vec{x})P_n\right)\, \varepsilon(\vec{x})=0\,.
\end{equation}
At each position $\vec{x}$, there are sixteen preserved supersymmetries by the same argument as before, but a supersymmetry $\varepsilon$ that is preserved globally has to be annihilated by all the projectors
\begin{equation} \label{projector_local_currents}
    \Pi(\vec{x})= \frac{1}{2}\left(1+\sum_{i=1}^n \frac{j_i(\vec{x})}{\mathcal{H}(\vec{x})}P_i\right) \,,
\end{equation}
for all $\vec{x}$.%
\footnote{Systems with sixteen local supersymmetries are called \textit{themelia} \cite{Bena:2022fzf}, and have been classified in \cite{Eckardt:2023nmn} (see also \cite{Eckardt:2024ugu}).}

We are interested in situations where some global  supersymmetries are preserved, so that the projector can be written as:
\begin{equation} \label{split_projector_mainbranes}
    \Pi(\vec{x})=f_1(\vec{x})\,\Pi_1+\ldots +f_m(\vec{x})\,\Pi_m \,,
\end{equation}
where the terms $\Pi_i$ and $f_i$ are $\Gamma$-matrix valued. The globally preserved supersymmetries are the ones in:
\begin{equation}
    \bigcap_{i=1}^m \ker \Pi_i \,.
\end{equation}
In this case, $\frac{32}{2^{m}}$ supersymmetries are preserved, with $m<n$.

In the context of black holes and of the Fuzzball programme, the globally preserved supersymmetries are the brane charges of the black hole. The other ones in \eqref{projector_local_currents} corresponds to dipolar excitations whose average value vanishes. Such dipolar excitations are sometimes called \textit{glues}. In other words, the supersymmetries of the black hole's brane charges are locally enhanced to sixteen by the presence of the glues. The logic, from the Fuzzball perspective, is that the smooth horizonless black-hole microstates carry dipolar excitations which are the fine-grained details of the geometry, while the classical black-hole solution only knows about the average value of its charges.

\subsection{Kappa-symmetry as a tool to probe branes} 
\label{sec:kappasym Non abelian}

The non-abelian brane probe action, $S_\mathrm{probe}$, is manifestly supersymmetric in the target space, but supersymmetry on the brane worldvolume is less evident. Indeed, one would at first sight find twice as many fermionic degrees of freedom on the brane than bosonic ones. This paradox is resolved by a projection condition that eliminates the unwanted fermionic degrees of freedom, thanks to the so-called $\kappa$ (kappa) symmetry, a fermionic symmetry of the full Born-Infeld action. All we need to know for our purposes is that  $\kappa$-symmetry is some local symmetry acting on the fermionic fields, and we can use it as a tool to probe the brane densities of a single brane in the Born-Infeld action \cite{Bena:2024oeq}. We describe the technical details thereof in the Appendix \ref{appendix:kappa_symm}. Reviews on $\kappa$-symmetry include \cite{Simon:2011rw,Bergshoeff:1997kr}. 

For our practical purposes, $\kappa$-symmetry results in a projection equation for the supersymmetry generator of the form:
\begin{equation}
    \left(\mathbbm{1}^{ab}-\Gamma_{\kappa}^{ab}\right)\varepsilon_{b}=0\,,
\end{equation}
where $a,b$ are gauge-group valued indices. This projection equation has a long-standing history in determining the number of global supersymmetries preserved by a solution. It has been shown in \cite{Bena:2024oeq} that, in the context of abelian gauge theories, this equation can be seen as a local projection equation and can be mapped to \eqref{projector_local} from which we can identify the brane densities. 
Although $\kappa$-symmetry has not been fully understood in the non-abelian version of Born-Infeld action, we still expect it to be a useful tool to determine local brane densities. In particular, it has been shown that the non-abelian version of $\kappa$-symmetry cannot be written as a $\mathrm{SU}(N)$ singlet: Physically, we interpret this fact as the impossibility to define a local supersymmetry structure in some contexts.

\vspace{1em}
In the remaining of the article, we study different examples and show that most of the brane configurations preserve sixteen local supersymmetries. In Section \ref{sec:monopoles}, we demonstrate this fact in the concrete example of the 't Hooft-Polyakov monopole, and show more generally that when the gauge group is maximally broken, the local supersymmetries behave exactly as if one has several different abelian branches, on which one can define the notion of local supersymmetry. In Section \ref{sec:brane_polarization}, we explain in detail the local supersymmetry structure when a D$(p+2)$-brane emerges from a D$p$-brane. In Section \ref{sec:instantons}, we show why instantons, describing D$p$-D$(p-4)$ bound states, do not have this sixteen-local-supersymmetry property, and explain that it sometimes arises from a perfect cancellation of the dipolar charges. In Section \ref{sec:3-charge_monopole}, we show that the three-charge version of the 't Hooft-Polyakov monopole with a momentum also preserves sixteen local supersymmetries.

\section{Monopoles}
\label{sec:monopoles}
In this section, we show that monopole solutions in the supersymmetric Yang-Mills theory (see \textit{e.g.} \cite{Manton:2004tk,Tong:2005un} for a review) have sixteen local supersymmetries. We start with the simple example of an $\mathrm{SU}(2)$ monopole: the so-called 't Hooft-Polyakov monopole. Then, we establish the result for general $\mathrm{U}(N)$ and $\mathrm{SU}(N)$ gauge groups. To do this, we compute the different brane densities using the Taylor-Van Raamsdonk method (see Section \ref{ssec:TVR_formalism}). We compute the densities directly from the super Yang-Mills action: the leading order of the full-fledged non-abelian Born-Infeld action. It is believed that BPS solutions are solutions of the full unknown non-abelian Born-Infeld action. Thus, the sixteen-local-supersymmetry feature of the brane densities that we are about to compute should extend to the full non-abelian Born-Infeld action as well.

\subsection{The 't Hooft-Polyakov monopole}
\label{ssec:tHooft-Polyakov}

First we study the most famous example of monopoles in super Yang-Mills (SYM) theory: the 't Hooft-Polyakov monopole. This is a solution of $\mathrm{SU}(2)$ super Yang-Mills in 3+1 dimensions, which can be seen as the theory living on the worldvolume of two D3 branes.

\subsubsection{The solution}
Monopoles are solutions to the bosonic action of the SYM theory in 4 dimensions,
\begin{equation} \label{SYM_action}
	S=-T_{D3}\int\mathrm{d}^4x  \tr\left( \frac{1}{4} (2\pi\alpha')^2(F_{ij}^a)^2+\frac{1}{2}(D_{\mu}\Phi^{a})^2 \right)\,,
\end{equation}
where $a$ is the gauge-group index. The field strength $F_{\mu\nu}$ and the covariant derivative are written as
\begin{equation}
	F_{\mu\nu}^{a} =\partial_{\mu}A_{\nu}^{a}-\partial_{\nu}A_{\mu}^{a}+\epsilon^{abc}A_{\mu}^bA_{\nu}^c \,, \qquad 
D_{\mu}\Phi^{a}=\partial_{\mu}\Phi^{a}+\epsilon^{abc}A_{\mu}^b\Phi^c\,.
\end{equation}
In particular, the 't Hooft-Polyakov monopole is a BPS solution given by \cite{Prasad:1975kr}:
\begin{subequations} \label{tHooftPolyakov_solution}
\begin{align} 
	&A_{i}^{a}=\epsilon_{aij}\frac{x^{j}}{r^2} \left(1-K(r)\right)\\
    &A_{0}^{a}=0 \\
	&\Phi^{a}=\frac{x^{a}}{r^2}H(r) \,, 
\end{align}
\end{subequations}
with 
\begin{equation}
K(r)=\frac{Pr}{\sinh(Pr)} \,, \qquad H(r)=Pr\coth(Pr)-1 \,,
\end{equation}
where $P$ is a constant that can be expressed in terms of the distance $\Delta Z$ separating the two D3 branes at infinity \cite{Hashimoto:1997px}.
Here, the radial coordinate is $r=\sqrt{(x^1)^2+(x^2)^2+(x^3)^2}$.\footnote{Note that in 3+1 dimensions, the gauge-group index $a=1,2,3$ can also be used to label the Cartesian coordinates in the three spatial dimensions, $x^a$, and that has been used in eq. \eqref{tHooftPolyakov_solution}.}

The solution \eqref{tHooftPolyakov_solution} solves a self-duality-like condition:
\begin{equation} \label{BPS-eq_tHP}
    D_i\Phi=\frac{1}{2}\epsilon_{ijk}(2\pi\alpha')F^{jk}\,.
\end{equation}
The variation of the fermions under supersymmetry has to be trivial for a bosonic background and is given by: 

\begin{equation}
    \delta \lambda=\left((D_i\Phi) \, \Gamma^{iz}+\frac{1}{2}(2\pi\alpha')F_{ij}\Gamma^{ij}\right)\varepsilon=0\,.
\end{equation}
Using the self-duality condition \eqref{BPS-eq_tHP}, the above equation becomes $\Gamma^{123z}\varepsilon=\varepsilon$.
This shows that this is a $\frac{1}{2}$-BPS solution of the SYM theory.\footnote{Note that here this is a $\frac{1}{2}$-BPS solution in the sense of super Yang-Mills.  But it is $\frac{1}{4}$-BPS with respect to the 32 supercharges of type-II string theory, as one has in addition the D3-brane projection equation.}

\begin{figure}[h]
    \centering
    \includegraphics[width=0.5\linewidth]{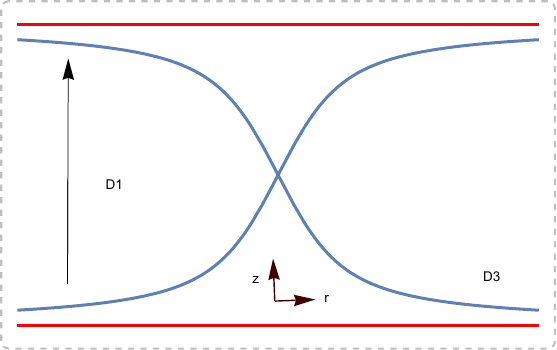}
    \caption{In blue, the shape taken by the D3-brane configuration induced by the scalar field $\Phi$ of a 't Hooft-Polyakov monopole. 
    There is a spherical symmetry in the directions $(x^1,x^2,x^3)$, and only the radius $r$ is represented here. The two D3-branes meet at $r=0$ and $z=0$. We represent in red the asymptotic separation of the two D3 branes. As we will show later, there exists a D1-brane current along the $z$ direction (represented by the black arrow), at all points in the region in between the two D3 branes. On the blue line there are densities of D1 and D3 branes.}
    \label{fig:enter-label}
\end{figure}

The presence of a non-trivial scalar field, $\Phi$, in the action \eqref{SYM_action} means that an extra-dimension, which we label $z$, with the rescaling $Z=2\pi\alpha'\Phi$, transverse to the worldvolume dimensions, is puffing up from the brane. After a gauge transformation, $Z=2\pi\alpha'\Phi$ can be diagonalized and its eigenvalues are
\begin{equation} \label{locus_D3branes_tHP}
z=\pm \pi \alpha' \frac{H(r)}{r} \equiv \pm \lambda(r) \,.
\end{equation}
The two branches of the above equation describes the location of two D3 branes that meet at the point where the radius $r$ shrinks to zero \cite{Hashimoto:1997px}, see Fig. \ref{fig:enter-label}. 

\subsubsection{The brane densities}

In the presence of a non-trivial scalar field (as in the configuration \eqref{tHooftPolyakov_solution}), an additional spatial direction transverse to the brane worldvolume, denoted by $z$, emerges. However, since the scalar field $\Phi$ does not commute with the gauge fields, extracting the brane-charge densities becomes more subtle. The technique introduced by Taylor and Hashimoto \cite{Hashimoto:2003pu,Hashimoto:2004fa}, is to compute the brane densities, $j_{\mathrm{D}p}(\vec{x})$, from the currents $J_{\mathrm{D}p}$ read off from the Taylor-Van~Raamsdonk formalism \eqref{Taylor-VRaamsdonk}. Formally it corresponds to first perform a Fourier transform in the extra ``non-abelian'' dimension, by inserting the operator $e^{i (2\pi \alpha')k_z \Phi}$ inside the symmetrised trace of $J_{\mathrm{D}p}$. From this, we obtain brane densities in Fourier space, $j_{\mathrm{D}p}(k_z)$. Finally, we perform the inverse Fourier transform of $j_{\mathrm{D}p}(k_z)$ and deduce the brane densities in terms of an abelian coordinate interpreted as the spatial position in the extra-dimension $z$. 

Note that the $e^{i (2\pi \alpha')k_z \Phi}$ insertion is essential.
Without this insertion, the brane densities, as directly read off from \eqref{non-commutative_WZ}, would have been:
\begin{subequations} 
\begin{align}
    &J_{\mathrm{D3}}^{0123} = ~T_{\mathrm{D3}}~\tr\left(\mathrm{I_d}\right) \\
    &J_{\mathrm{D3}}^{0zij} =~~T_{\mathrm{D3}} ~\epsilon_{ijk}\tr\left( D_{k}\Phi\right) =0\\
    &J_{\mathrm{D1}}^{0i} ~=~~ T_{\mathrm{D3}}(2\pi\alpha')\frac{\epsilon_{ijk}}{2}\tr\left(F_{jk}\right)=0\\
    &J_{\mathrm{D1}}^{0z}~ = ~~T_{\mathrm{D3}}~(2\pi\alpha')\frac{\epsilon_{ijk}}{2}\tr\left( F_{ij} D_{k}\Phi\right)    \,,
\end{align}
\end{subequations}
and the vanishing of the two equations in the middle would have led to the conclusion that monopoles do not preserved sixteen local supercharges. It would look like that because $\Phi$ and $F_{ij}$ are traceless, the ``glues/dipoles vanish". However, to reveal the dipole charges, the trick is thus to perform a Fourier transform at operator level (\textit{i.e.} inserting the $e^{i (2\pi \alpha')k_z \Phi}$ in the trace), and inverse Fourier transform back.%
\footnote{This trick is in fact similar to Myers' technique to reveal the D2-brane density out of the D0-brane worldline \cite{Myers:1999ps}: inside the trace was performed a `non-Abelian' Taylor expansion in which a spatial coordinate is replaced by its corresponding operator.}

So, after inserting the $e^{i (2\pi \alpha')k_z \Phi}$ in the trace, we deduce the brane densities in Fourier space \cite{Hashimoto:2003pu,Hashimoto:2004fa}:
\begin{subequations} \label{D3D1densities_tHooftPolyakov}
\begin{align}
    &\tilde{j}_{\mathrm{D3}}^{0123}(k_z,r) = T_{\mathrm{D3}}\str\left(e^{i (2\pi \alpha')k_z \Phi}\right) \\
    &\tilde{j}_{\mathrm{D3}}^{0zij}(k_z,r) =T_{\mathrm{D3}} \epsilon_{ijk}\str\left(e^{i (2\pi \alpha')k_z \Phi} D_{k}\Phi\right) \label{D1D3-glueD3} \\
&\tilde{j}_{\mathrm{D1}}^{0i}(k_z,r) = T_{\mathrm{D3}}(2\pi\alpha')\frac{\epsilon_{ijk}}{2}\str\left(e^{i (2\pi \alpha')k_z \Phi} F_{jk}\right) \label{D1D3-glueD1}\\
&\tilde{j}_{\mathrm{D1}}^{0z}(k_z,r) = T_{\mathrm{D3}}(2\pi\alpha')\frac{\epsilon_{ijk}}{2}\str\left(e^{i (2\pi \alpha')k_z \Phi} F_{ij} D_{k}\Phi\right)  \label{eq:D1 brane density}  \,,
\end{align}
\end{subequations}
with the indices $i,j,k \in \{1,2,3\}$ labelling the three Cartesian spatial coordinates.

The BPS equation \eqref{BPS-eq_tHP}, relating $F$ to $D_i\Phi$, ensures that the densities of the glues, \eqref{D1D3-glueD3} and \eqref{D1D3-glueD1}, are equal. This implies that the projection condition, in Fourier space, is of the form \eqref{projector_local}, ensuring the presence of sixteen supersymmetries. 

However, the local supersymmetries in position space are more involved. After plugging in the expression of the monopole solution we obtain:
\begin{subequations}
\begin{align}
    &\tilde{j}_{\mathrm{D3}}^{0123}(k_z,r) = T_{\mathrm{D3}}\left(e^{i k_z \lambda(r)}+e^{-i k_z \lambda(r)}\right) \\
    &\tilde{j}_{\mathrm{D3}}^{0zij}(k_z,r) =T_{\mathrm{D3}}(2\pi\alpha') \epsilon_{ijk}\frac{x^{k}}{2 r^3}(1-K^2)\left(e^{i k_z \lambda(r)}-e^{-i k_z \lambda(r)}\right) \\
    &\tilde{j}_{\mathrm{D1}}^{0i}(k_z,r) =\frac{\epsilon_{ijk}}{2}\tilde{j}_{\mathrm{D3}}^{0zjk}(k_z)\\
    &\tilde{j}_{\mathrm{D1}}^{0z}(k_z,r) = T_{\mathrm{D3}}(2\pi\alpha')^2\frac{1}{4 r^4}(1-K^2)^2\left(e^{i k_z \lambda(r)}+e^{-i k_z \lambda(r)}\right) \nonumber\\
    &\hspace{8em} +\frac{1}{i(2\pi\alpha')k_z}\frac{H K^2}{2\pi \alpha' r^3}\left(e^{i k_z \lambda(r)}-e^{-i k_z \lambda(r)}\right) \label{eq: monopole SU(2) off diagonal}  \,.
\end{align}
\end{subequations}
Then the inverse Fourier transform gives the brane densities in position space \cite{Hashimoto:2003pu,Hashimoto:2004fa}: 
\begin{subequations} \label{D3-D1-densities_position_tHP}
\begin{align}
    j_{\mathrm{D3}}^{0123}(z,r)=~& T_{\mathrm{D3}} \left[ \delta\left(z-\lambda(r)\right)+\delta\left(z+\lambda(r)\right)\right]\\
    j_{\mathrm{D3}}^{0zij}(z,r)=~& T_{\mathrm{D3}}(2\pi\alpha') \epsilon_{ijk}\frac{x^{k}}{2 r^3}(1-K^2) \left[ \delta\left(z-\lambda(r)\right)-\delta\left(z+\lambda(r)\right)\right] \\
    j_{\mathrm{D1}}^{0i}(z,r)=~& T_{\mathrm{D3}}(2\pi\alpha')\frac{x^{i}}{2 r^3}(1-K^2) \left[ \delta\left(z-\lambda(r)\right)-\delta\left(z+\lambda(r)\right)\right]\\
    j_{\mathrm{D1}}^{0z}(z,r)=~ & T_{\mathrm{D3}}(2\pi\alpha')^2 \biggl[\frac{1}{4 r^4}(1-K^2)^2\left( \delta\left(z-\lambda(r)\right)+\delta\left(z+\lambda(r)\right)\right) \nonumber\\
    &\hspace{7em} +\frac{H K^2}{2\pi \alpha' r^3}\left( \Theta\left(z-\lambda(r)\right)-\Theta\left(z+\lambda(r)\right)\right)\biggr] \,, \label{j_D1_main_DiracHeaviside}
\end{align}
\end{subequations}
where $\delta$ and $\Theta$ denote respectively the Dirac and Heaviside distributions.

\subsubsection{The brane projector}

Before studying the local supersymmetries, we first describe the solution of equations \eqref{D3-D1-densities_position_tHP}.
We observe the existence of D3- and D1-brane densities at the locus \eqref{locus_D3branes_tHP} described by the scalar $\Phi$, through the Dirac distributions, $\delta$. In addition, there is also a local D1-brane density between the two D3-branes given by the Heaviside distributions, $\Theta$. Therefore, there are three relevant regions of space to consider:
\begin{itemize}
    \item The upper branch, parametrized by $\delta\left(z-\lambda(r)\right)$\,.
    \item The lower branch, parametrized by $\delta\left(z+\lambda(r)\right)$\,.
    \item The space in between the two D3 branes characterized by $\Theta\left(z-\lambda\right)-\Theta\left(z+\lambda\right)$\,.
\end{itemize}

The Hamiltonian density is given by $-\mathcal{L}$, the Lagrangian evaluated on-shell, from the worldvolume theory: 
\begin{equation}
    \mathbb{H}=T_{D3}\int r^2\mathrm{d}r\mathrm{d}\Omega_2 \left(  \tr\left( 1+\frac{1}{4} (2\pi\alpha')^2(F_{\mu \nu}^a)^2+\frac{1}{2}(D_{\mu}\Phi^{a})^2 \right)\right)\,.
\end{equation}
Similar to the computation of the brane densities above, the Hamiltonian density naturally splits in two parts, on top of the branes and in between the branes: 
\begin{equation}
    \tilde{\mathcal{H}}(z,r)= \mathcal{H}_{\mathrm{on top}}(r) \left[\delta\left(z-\lambda\right) + \delta\left(z-\lambda\right) \right]
    + \tilde{\mathcal{H}}_{\mathrm{btw}}(z,r) \left[ \Theta\left(z-\lambda\right)-\Theta\left(z+\lambda\right) \right] \,.
\end{equation}

\paragraph{Upper branch.}
On the upper and lower branches, the energy density on the worldvolume of the brane simplifies to 
\begin{equation}
\mathcal{H}_{\mathrm{on top}}(r)
= T_{\mathrm{D3}} \left( 1+ \chi(r)^2 \right) \,,
\end{equation}
where we have defined $\chi(r) = 2\pi\alpha'\frac{(1-K(r)^2)}{2 r^2}$. 

On the upper branch, the projection equation reads
\begin{equation} \label{projector_ontop_upper}
    \left(1+\frac{1}{1+\chi^2}P_{\mathrm{D3},\,r}+\frac{\chi}{1+\chi^2}P_{\mathrm{D3},\,z}+\frac{\chi}{1+\chi^2}P_{\mathrm{D1},\,r}+\frac{\chi^2}{1+\chi^2}P_{\mathrm{D1},\,z}\right)\varepsilon=0\,,
\end{equation}
and shows that there are sixteen supersymmetries locally at the location of the brane. 
Besides, one can express the above expression \eqref{projector_ontop_upper} in terms of the tilt of the D3-brane shape \eqref{locus_D3branes_tHP},  $\tan \alpha \equiv \frac{\pd z}{\pd r}$, since calculations show that $\frac{\pd z}{\pd r}= \chi(r)$. In other words, the tilt of the brane shape determines the brane density ratios.

\paragraph{Lower branch.}
Similarly, on the lower branch, we have:
\begin{equation}
    \left(1+\frac{1}{1+\chi^2}P_{\mathrm{D3},\,r}-\frac{\chi}{1+\chi^2}P_{\mathrm{D3},\,z}-\frac{\chi}{1+\chi^2}P_{\mathrm{D1},\,r}+\frac{\chi^2}{1+\chi^2}P_{\mathrm{D1},\,z}\right)\varepsilon=0\,,
\end{equation}
which shows that there are also sixteen local supersymmetries on the lower branch.
With a BIon solution \cite{Callan:1997kz} describing an infinite D3-D1 spike, one would have obtained only one of the two branches. The 't Hooft-Polyakov solution, which describes two spikes joining at a point, contains two different branches of the `abelian' projection equation, each with opposite signs in the dipolar charges. 

\begin{figure}[h!]
    \centering
    \includegraphics[scale=1]{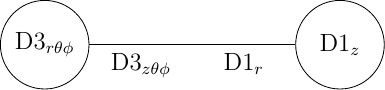}
    \caption{Diagram of the brane densities on top of the branes. The global charges, whose charges do not vanish when summed over both branches of the D3 locus, are encircled. The `glues', whose charges are opposite on either branch, are represented below the line linking the `main branes'.} 
    \label{fig:D3-D1_diagram}
\end{figure}

It is convenient to represent the brane densities in a diagram, see Figure \ref{fig:D3-D1_diagram}.

\paragraph{Between the branes.}
Between the two D3-branes, we only have D1-brane flux along $z$. The D1-brane density, which is also the energy density in between the branes, is  $\tilde{\mathcal{H}}_{\mathrm{btw}}(z,r)=T_{\mathrm{D3}}(2\pi\alpha')\frac{H K^2}{r^3}$; after integrating over $z$, we have
\begin{equation}
\mathcal{H}_{\mathrm{btw}}(r)=T_{\mathrm{D3}}\, 2(2\pi\alpha')^2\left( \frac{H(r)K(r)}{r^2} \right)^2 \,.
\end{equation}

The projection equation in the region between the branes is that of a D1 brane alone along $z$
\begin{equation}
\left(1+P_{\mathrm{D1},\,z}\right)\varepsilon=0\,,
\end{equation}
and this ensures sixteen local supersymmetries in this region.

\subsection{Monopoles with general non-abelian gauge group}
\label{ssec:monopoles_general}

In this section, we consider a general monopole of $\mathrm{U}(N)$ or $\mathrm{SU}(N)$ super Yang-Mills in $3+1$ dimensions \cite{Diaconescu:1996rk}.
Monopoles are solutions of the equations
\begin{equation}
    (2\pi\alpha')\frac{\epsilon_{ijk}}{2}F_{jk}= D_i\Phi\,.
\end{equation}
We aim at computing the brane densities \eqref{D3D1densities_tHooftPolyakov} for such monopole solutions. 

\paragraph{Density of D3(123).}
By use of gauge transformations, the scalar field $\Phi$ can be diagonalized and therefore takes only values in the Cartan sub-algebra: $\Phi={\rm diag}\left(\phi_1,\phi_2,\dots,\phi_N\right)$.\footnote{For the purposes of this computation, we treat $\mathrm{SU}(N)$ and $\mathrm{U}(N)$ monopoles equivalently, as the computations are identical. The only distinction at this stage is whether the field $\Phi$ has zero trace or not.} 
Thus, the D3-brane current along the directions 123 is
\begin{align}
    &\tilde{j}_{\mathrm{D3}}^{0123}(k_z)= T_{\mathrm{D3}}\,\mathrm{STr}\left(e^{i (2\pi \alpha')k_z \Phi}\right)
    = T_{\mathrm{D3}} \sum_{a=1}^N e^{i (2\pi \alpha')k_z \phi_a} \,.
\end{align}

\paragraph{Density of the glues.}
Next, we compute the currents of the glues, namely $\tilde{j}_{\mathrm{D3}}^{0zij}(k_z)$ and $\tilde{j}_{\mathrm{D1}}^{0i}(k_z)$ in \eqref{D3D1densities_tHooftPolyakov}.

The D3 density $\tilde{j}_{\mathrm{D3}}^{0zij}(k_z)$ involves the expression $D_{i}\Phi=\partial_i\Phi+[A_i,\Phi]$. 
In order to perform the computation, we decompose $\Phi$ and $A_k$ in the `canonical' basis for $N\times N$ matrices. This basis consists of the matrices $(E_{ab})_{a,b=1,\dots,N}$, whose only non-vanishing component is a $1$ at the position $(a,b)$ in the matrix: they correspond to the operators $\ket{e_a}\bra{e_b}$ in the canonical basis of vectors.

Then a simple computation shows that the diagonal components of
\begin{equation}
\label{eq:Commutator covariant derivative}
    [A_i, \Phi]=\sum_{a,b=1}^N(\phi_b-\phi_a)A_{i}^{ab}E_{ab} 
\end{equation}
vanish. Therefore, the contribution ${\rm Tr}\left(e^{i(2\pi\alpha')k_z \Phi} [A_{i},\Phi]\right)$ vanishes, since $e^{i(2\pi\alpha')k_z \Phi}$ is a diagonal matrix.
Thus, we obtain:
\begin{equation}
    {\rm Tr}\left( e^{i (2\pi \alpha')k_z \Phi}D_{i}\Phi\right) 
    =\frac{1}{i (2\pi \alpha')k_z}\partial_{i} \left[ {\rm Tr}\left( e^{i (2\pi \alpha')k_z \Phi}\right) \right]\,.
\end{equation}

It is consequently straightforward to compute the D3-glue density:
\begin{align}
    j_{\mathrm{D3}}^{0zij}(k_z) = T_{\mathrm{D3}}(2\pi \alpha')\epsilon^{ijk} \left( e^{i (2\pi \alpha')k_z \phi_1} \partial_k \phi_1 +\dots + e^{i (2\pi \alpha')k_z \phi_N} \partial_k\phi_N \right) \,.
\end{align}
And using the BPS equations, one deduces the D1-glue density:
\begin{align}
    j_{\mathrm{D1}}^{0k}(k_z)=\frac{\epsilon^{kij}}{2}j_{{\rm D3}}^{0zij}(k_z)= T_{\mathrm{D3}}(2\pi \alpha') \left( e^{i (2\pi \alpha')k_z \phi_1} \partial_k \phi_1 +\dots + e^{i (2\pi \alpha')k_z \phi_N} \partial_k\phi_N \right) \,.
\end{align}

\paragraph{Density of D1($z$).}
We are left with computing the D1-brane density along $z$. By definition and by the BPS equations, this is given by:
\begin{equation}
    j_{\mathrm{D1}}^{0z}(k_z)=T_{\mathrm{D3}}(2\pi \alpha') \frac{\epsilon^{ijk} }{2}\mathrm{STr}\left(e^{i (2\pi \alpha')k_z \Phi}F_{ij}D_{k}\Phi\right)=T_{\mathrm{D3}}~\mathrm{STr}\left(e^{i (2\pi \alpha')k_z \Phi}D^{k}\Phi D_{k}\Phi\right)\,.
\end{equation}
The computation is quite involved, so we defer the detailed derivation to Appendix \ref{sec:appendix:brane densities general monopole}.
In short, the total current is given by:
\begin{equation}
    j_{\mathrm{D1}}^{0z}(k_z)=T_{\mathrm{D3}} \left(\sum_{a=1}^N\left(\partial_i \phi_a\right)^2 e^{i (2\pi \alpha')k_z \phi_a} + \sum_{a,b=1}^N\frac{2(\phi_a-\phi_b)}{i 2\pi \alpha'k_z}e^{i(2\pi \alpha')k_z \phi_a}A_{i}^{ab}A^{i,\,ba}\right)\,.
\end{equation}
After inverse Fourier transform, the first and second term of the above current yield the Dirac and Heaviside distributions respectively, like in \eqref{j_D1_main_DiracHeaviside}.
The brane densities may not seem gauge invariant, but this is because we chose a specific gauge: the one which diagonalizes the scalar field.

A direct application of the above formula is to use it to recover the  densities of the $\mathrm{SU}(2)$ monopole. One must be careful, as the normalization is slightly different from the usual decomposition in terms of the normal basis of generators of the Lie algebra. For an $\mathrm{SU}(2)$ monopole, in the gauge where the scalar field is diagonal (hence $\phi_1 = -\phi_2 \equiv\phi$), the off-diagonal contribution is given by:
\begin{equation}
    j_{\text{off-diag}}= \frac{4\phi}{i2\pi \alpha'k_z } A_i^{12}A^{i,\,21}\left(e^{i 2\pi \alpha' k_z \phi}-e^{-i 2\pi \alpha' k_z \phi}\right)\,.
\end{equation}
We recover the results of the previous section. In particular, this term is equal to the second line of \eqref{eq: monopole SU(2) off diagonal}.

\paragraph{Brane densities in coordinate space.}

Performing the Fourier transform, we obtain all the currents as follow:
\begin{subequations} \label{eq:monopole final}
\begin{align}
    &j_{\mathrm{D3}}^{0ijk}(z)=T_{\mathrm{D3}} \sum_{a=1}^N \delta\left(z-\phi_a\right)\\ 
    &j_{\mathrm{D3}}^{0zij}(z)~=~T_{\mathrm{D3}}~\epsilon^{ijk} ~\sum_{a=1}^N\partial_k \phi_a \delta\,\left(z-\phi_a\right) \\ 
    &j_{\mathrm{D1}}^{0k}(z)~=~\frac{\epsilon^{kij}}{2}~j_{{\rm D3}}^{0zij} \\
    &j_{\mathrm{D1}}^{0z}(z)~=~T_{\mathrm{D3}} ~\left( \sum_{a=1}^N\partial_i \phi_a \partial^i\phi_a \delta\left(z-\phi_a\right)+\sum_{a,b=1}^N 2(\phi_a-\phi_b) A_{i}^{ab}A^{i,\,ab} \Theta\left(z-\phi_a\right)\right)\,. \label{eq:monopole general in coordinate space}
\end{align}
\end{subequations}
Similar to the 't Hooft-Polyakov case, the brane currents can be split between the region away from the branes and the locus of the branes.

\paragraph{Away from the branes.}
We called $z$ the coordinate which labels the extra-direction puffing out of the branes. Away from the branes, all the Dirac distributions of \eqref{eq:monopole final} vanish. The last expression shows that beyond the region delimited by the largest and smallest eigenvalues of the scalar field, the Heaviside functions $\Theta$ cancel each other and, as expected for such a monopole, there is no D1 brane density outside the worldvolume of the branes. The interpretation of the second term of \eqref{eq:monopole general in coordinate space} is that of open strings stretching between the branes labelled by $a$ and $b$, and which become massive. Indeed, integrating over $z$ between the $a^\textbf{th}$ brane and the $b^\textbf{th}$ brane yields the terms $(\phi_a-\phi_b)^2 A_{i}^{ab}A^{i,\,ab}$: This is exactly the mass of the stretched string times the norm squared of $A$. Thus, we learn that the massive open-string excitations give rise to D1-brane densities outside of the brane worldvolume.

\paragraph{On top of the branes.}
On top of the brane labelled by $a$, \eqref{eq:monopole final} becomes
\begin{subequations}
\begin{align}
    &j_{\mathrm{D3}}^{0ijk}=T_{\mathrm{D3}} \\
    &j_{\mathrm{D3}}^{0zij}=T_{\mathrm{D3}} (2\pi \alpha') \epsilon^{ijk} \partial_k \phi_a\\ 
    &j_{\mathrm{D1}}^{0k}=\frac{\epsilon^{kij}}{2}j_{{\rm D3}}^{0zij}\\ 
    &j_{\mathrm{D1}}^{0z}=T_{\mathrm{D3}} (2\pi \alpha')^2 \left(\partial_i \phi_a \partial^i\phi_a \right)\,.
\end{align}
\end{subequations}
The density of energy on the $a^\mathrm{th}$ brane is $\mathcal{H}=T_{\mathrm{D3}}\left(1+(\partial_i \phi_a)^2 \right)$, and we obtain the projection equation
\begin{equation}
    \left(1+\frac{\epsilon^{ijk}}{6}\frac{P_{\mathrm{D3},\,ijk}}{1+(\partial_\ell\phi_a)^2}
    +\frac{\epsilon^{ijk}}{2}\frac{\partial_k\phi_a\,P_{\mathrm{D3},\,ijz}}{1+(\partial_\ell\phi_a)^2} 
    +\frac{\partial_i\phi_a \,P_{\mathrm{D1},\,i}}{1+(\partial_\ell\phi_a)^2}
    +\frac{(\partial_\ell\phi_a)^2\,P_{\mathrm{D1},\,z}}{1+(\partial_\ell\phi_a)^2}\right)\varepsilon=0\,,
\end{equation}
which shows the sixteen local supersymmetries.

\vspace{1em}
In this Subsection, we showed that $\mathrm{SU}(N)$ monopoles preserve sixteen local supersymmetries, on top of the D3-brane worldvolume and the region away from them. Besides, we noted that the D1-brane densities outside of the D3-brane worldvolume comes from the open-string excitations which become massive.

\section{Brane polarization}
\label{sec:brane_polarization}

In this section we study another class of geometries of interest: low dimensional branes that polarize into higher dimensional one. We consider Myers' solution of D0-branes, which polarize into D2-branes due to an external field strength \cite{Myers:1999ps}, and show that this kind of solution has also sixteen local supersymmetries. This result is straightforward to generalize to other instances, such as the polarization of D0-branes in the context of the BMN matrix model\cite{Berenstein:2002jq}, in the non-commutative BIon \cite{Constable:1999ac}, or, for example, when masses are given to the chiral multiplets of the $\mathcal{N}=1$ formulation of $\mathcal{N}=4$ super Yang-Mills, resulting in D3-branes that polarize into five-branes \cite{Polchinski:2000uf}.

In contrast to the abelian Born-Infeld action, a stack of branes can couple to higher-dimensional branes; in particular, a stack of D0-branes can couple to a D2-brane, as explained in Section \ref{sec:brane density formalism}. Consider a stack of $N$ D0-branes, described by the maximally supersymmetric (0+1)-dimensional $\mathrm{SU}(N)$ super Yang-Mills living on the worldvolume of the D0-branes.
Under the action of an external four-form field strength, $F_{0123}$, the bosonic part of the action, after some transformations on the Chern-Simons term, is given by \cite{Myers:1999ps}:
\begin{equation}
    S=-T_{\mathrm{D0}} \int {\rm d}t \,{\rm Tr}\left(\frac{1}{2}\partial_0\Phi_i\partial^0\Phi^i+\frac{1}{4}[\Phi_i,\Phi_j][\Phi^j,\Phi^i]\right)-iT_{\mathrm{D0}}\int {\rm d}t \,\frac{F_{0ijk}}{3} \,{\rm Tr}\left(\Phi^i\Phi^j\Phi^k\right)\,.
\end{equation}
The indices $i,j,k=1,2,3$ label three of the nine scalar fields $\Phi_{i=1,2,3}$ that couple to the four-form field strength $F$. 
As such, three spatial dimensions emerge out of the D0-brane worldvolume theory.

We consider a spherically symmetric background field strength:
\begin{equation}
    F_{0ijk}=-2 f\epsilon_{ijk}\,,
\end{equation}
and the equations of motion for the static solutions simplify to:
\begin{equation}
  [ [\Phi_i,\Phi_j],\Phi^j]+i f \epsilon_{ijk}[\Phi^j,\Phi^k] =0 \,.
\end{equation}
The trivial configuration satisfies the equation of motion but is a maximum. A minimum is obtained when we consider the solution:
\begin{equation} \label{fuzzy_S2}
    \Phi_{i}=\frac{f}{2}L_i \,,
\end{equation}
where the $L_i$ form an $N$-dimensional representation of $\mathrm{su}(2)$, satisfying $[L_i,L_j]=2i\epsilon_{ijk}L_k$. From now on, we work only with irreducible representation of dimension $N$ and the results will be straightforward to generalize to any representation. The configuration described by the D0-brane matrices $\Phi^i$ \eqref{fuzzy_S2} is called a fuzzy two-sphere \cite{Kabat:1997im}. 
The question we want to address in this section is: Does this solution have sixteen local supercharges? 

From the Taylor-Van Raamsdonk formalism, the currents for D0 and D2 branes are:
\begin{equation}
    \tilde{j}_{\mathrm{D0}}^0(\vec{k})=T_{\rm D0}\str\left(e^{i \vec{k} \cdot\vec\Phi}\right)\,,\quad 
    \tilde{j}_{\mathrm{D2}}^{0ij}(\vec{k})=\frac{T_{\rm D0}}{2\pi\alpha'}\str\left(-ie^{i \vec{k} \cdot\vec\Phi}[\Phi_i,\Phi_j]\right)\,.
\end{equation}
We focus on irreducible representation of $\mathrm{su}(2)$ of dimension $N$, and we obtain the density of charges:
\begin{equation}
\label{eq:density of brane polarize momentum}
    \tilde{j}_{\mathrm{D0}}^0(\vec{k})=T_{\rm D0}N \left(e^{i|k| R}+e^{-i |k| R}\right)\,,\quad \tilde{j}_{\mathrm{D2}}^{0ij}(\vec{k})=\frac{T_{\rm D0}N}{2\pi\alpha'}\epsilon_{ijl}\frac{k_l}{|k|}f R\left(e^{i|k| R}-e^{-i|k| R}\right)\,,
\end{equation}
where $|k|=\sqrt{\sum_i (k^i)^2}$ is the norm of $\vec{k}$ and $R^2=f^2 \left(N^2-1\right)$.%
\footnote{The first expression of \eqref{eq:density of brane polarize momentum} is straightforward to compute. The second one can be computed by noting that $\str\left(-ie^{i \vec{k}\cdot \vec\Phi}[\Phi_i,\Phi_j]\right)=f \epsilon_{ijl}\str\left(e^{i \vec{k}\cdot \vec\Phi}\Phi_l\right)=\frac{f \epsilon_{ijl}}{i}\partial_{k^l} \str\left(e^{i \vec{k}\cdot \vec\Phi}\right)$, and the result follows.}
Performing the inverse Fourier transform of the above quantities, we obtain in spherical coordinates:
\begin{equation}
\label{eq:density of brane polarize}
    j_{\mathrm{D0}}^0(r)=\frac{T_{\rm D0}N}{4\pi R}\left(\frac{1}{R}\delta(r-R)+\delta'(r-R)\right)\,,\quad 
    j_{\mathrm{D2}}^{0ij}(r)=\frac{T_{\rm D0}N}{2\pi\alpha'} f R\frac{\epsilon_{ijl}\,x^{l}}{4\pi R\,r}\,\delta'(r-R) \,.
\end{equation}

Perhaps an easier way to extract physics from the densities of branes \eqref{eq:density of brane polarize}, is to break momentarily the spherical symmetry of the result. Pick one specific point in Fourier space, $\vec{k}=(0,0,k_3)$. Then \eqref{eq:density of brane polarize momentum} becomes:
\begin{equation}
    \tilde{j}_{\mathrm{D0}}^0(k_3)=T_{\mathrm{D0}} N\left(e^{ik_3R}+e^{-ik_3R}\right)\,,\quad
    \tilde{j}_{\mathrm{D2}}^{012}(k_3)=\frac{T_{\mathrm{D0}}N}{2\pi\alpha'}f R\left(e^{ik_3R}-e^{-ik_3R}\right)\,,
\end{equation}
and after Fourier transform, this gives:
\begin{equation}
    j_{\mathrm{D0}}^0(x_3)=T_{\mathrm{D0}} N\left(\delta(x_3-R)+\delta(x_3+R)\right)\,,\quad 
    j_{\mathrm{D2}}^{012}(x_3)=\frac{T_{\rm D0}N}{2\pi\alpha'} f R \left(\delta(x_3-R)-\delta(x_3+R)\right)\,.
\end{equation}
This expression describes the D0-brane and D2-brane densities at the `north' pole ($x_3=R$) and `south' pole ($x_3=-R$) of the $\mathbb{S}^2$. 
The D0-brane densities on the two poles adds up, while the D2-brane densities have opposite signs and cancel each other. As such, the D2-brane charges are dipolar: there is no net D2-brane charge on the entire sphere. 
Restoring the spherical symmetry, along a fixed unit vector $\vec{u}$, the brane densities become
\begin{subequations}
\begin{align}    
    j_{\mathrm{D0}}^0(r\vec{u})&=T_{\mathrm{D0}} N\, \delta(r-R)\,,\\
    j_{\mathrm{D2}}^{0ij}(r\vec{u})&=\frac{T_{\rm D0}N}{2\pi\alpha'} f R \epsilon_{ijk} u^k\,\delta(r-R)\,.
    \label{eq:density of brane polarize2}
\end{align}
\end{subequations}
The D2-brane density at any point $r\vec{u}$ is exactly cancelled by that of its antipodal point $-r\vec{u}$: there is no net D2-brane charge when one integrates over the $\mathbb{S}^2$. Note that \eqref{eq:density of brane polarize} and \eqref{eq:density of brane polarize2} differ because the former involve a three-dimensional $\delta$-function while the latter involve a one-dimensional one.

At the `north' pole, at $(x_1,x_2,x_3)=(0,0,R)$, the projection equation is: 
\begin{equation}
    \left(1+\frac{\Gamma^{0}i\sigma_2+\frac{f R}{2\pi\alpha'}\Gamma^{012}\sigma_1}{\sqrt{1+\left(\frac{f R}{2\pi\alpha'}\right)^2}}\right)\varepsilon=0\,.
\end{equation}
And at a generic position $\vec{x}$ on the sphere of radius $R$, the projection equation is:
\begin{equation}
    \left(1+\frac{\Gamma^{0}i\sigma_2+\frac{f \epsilon_{ijk}x^k}{4\pi\alpha'}\Gamma^{0ij}\sigma_1}{\sqrt{1+\left(\frac{fR}{2\pi\alpha'}\right)^2}}\right)\varepsilon=0\,.
\end{equation}
Thus, there are sixteen local supersymmetries at all points on the fuzzy two-sphere.

\vspace{1em}
It is easy to generalise the computation to any representation of $\mathrm{su}(2)$. 
A generic $N$-dimensional representation of $\mathrm{su}(2)$ can be decomposed into a direct sum of irreducible representation of dimension $n_k$ such that $\sum_k n_k= N$. The currents \eqref{eq:density of brane polarize} are naturally replaced by:
\begin{align}
    j_{\mathrm{D0}}^0(r)&=T_{\rm D0}~\sum_k ~\frac{n_k}{4\pi R_k}\left(\frac{1}{R_k}\delta(r-R_k)+\delta'(r-R_k)\right)\,,\\ 
    j_{\mathrm{D2}}^{0ij}(r)&=T_{\rm D0}~\sum_k ~\frac{n_k}{2\pi\alpha'} f R_k\frac{\epsilon_{ijl}\,x^{l}}{4\pi R_k\,r}\,\delta'(r-R_k) \,,
\end{align}
with $R_k^2=f^2 \left(n_k^2-1\right)$.

Moreover, this computation goes beyond the case of a D0 brane polarizing into a D2 brane, and generalises for other well-known brane polarization effects.
For example, for mass deformations of $\mathcal{N}=4$ super Yang-Mills \cite{Polchinski:2000uf}, each $\mathcal{N}=1^{*}$ vacua are labelled by $\mathrm{su}(2)$ representations and corresponds to D3 branes polarizing into D5 branes and the brane density computation would be rather similar to what was computed in this section. The same computation can also be performed for all the mass-deformed maximally supersymmetric Yang-Mills theory in $d+1$ dimensions, which corresponds to D$p$ branes polarizing into D$(p+2)$ branes.

\section{Instantons}
\label{sec:instantons}

In this section, we consider instantons of maximally supersymmetric Yang-Mills (SYM) theories in $4+1$ dimensions, described by the action:
\begin{equation}
    S= -\frac{1}{4}T_{D4}(2\pi\alpha')^2\int\,d^5x\,\tr{F_{\mu\nu}\,F^{\mu\nu}}\,.
\end{equation}
We did not write all the fermions and scalar fields since they will be trivial; the fact that we use maximally supersymmetric Yang-Mills rather than pure Yang-Mills theory is only used at the level of the supersymmetry variation.

Instantons are finite-action solutions of the equations of motion, and satisfy the self-duality equation:
\begin{equation}
\label{eq:self-duality}
    F_{ij}=\star F_{ij}\,,
\end{equation}
where $i,j$ label the 4d spatial coordinates.
Importantly, instanton solutions in $p$-dimensional SYM can be viewed as describing $D(p-4)$ branes inside the worldvolume of D$p$-branes. Reviews on SYM instantons include \cite{Belitsky:2000ws,Manton:2004tk,Tong:2005un,Dorey:2002ik}. In the following, we focus on D0-D4 instantons, but the discussion can be generalized to other D$(p-4)$-D$p$ configurations. 

In this Section, we show that SYM instantons do not have sixteen local supercharges. However, we show that in some cases, the absence of sixteen local supersymmetries results from a cancellation between the local dipole charges that can be revealed by going infinitesimally into the Coulomb branch.

\subsection{A toy model: the smeared instanton}
\label{sec:Smeared instanton}

We first consider a simple toy model: the smeared instanton of $\mathrm{SU}(2)$ Yang-Mills. On the worldvolume of the two D4 branes, we consider a constant field strength $F$ along the Pauli matrix $\sigma_3$: 
\begin{equation}
    F=F_{12} \,\sigma_3 \, dx^1 \wedge dx^2 + F_{34}\,\sigma_3 \, dx^3 \wedge dx^4 \,,
\end{equation}
such that $F_{12}=F_{34}$.
The configuration preserves eight supersymmetries and satisfies the self-duality condition \eqref{eq:self-duality}. 

Since the integrand of the Wess-Zumino action \eqref{non-commutative_WZ} is of the form $C^{(5)} + (2\pi\alpha')C^{(3)}\wedge F + \frac{1}{2}(\twopia)^2 C^{(1)}\wedge F\wedge F$, one easily deduces that the brane densities are given by
\begin{subequations} 
\begin{align}
&J_{\mathrm{D4}}^{01234}~=~T_{\mathrm{D4}}\tr(\mathbbm{1}) \\
    &J_{\mathrm{D2}}^{012}~~~=~T_{\mathrm{D4}} (\twopia) \tr(F_{34}\sigma_3)\\
    &J_{\mathrm{D2}}^{034}~~~=~T_{\mathrm{D4}} (\twopia) \tr(F_{12}\sigma_3)\\
    &J_{\mathrm{D0}}^0~~~~=~ T_{\mathrm{D4}} (\twopia)^2 \tr (F_{12}F_{34}(\sigma_3)^2)\,.
\end{align}
\end{subequations}
Because $\mathrm{Tr}\left(\sigma_3\right)=0$, the density of D2 branes vanish and we have:
\begin{subequations} \label{D0-D4_smeared}
\begin{align}
    J_{\mathrm{D4}}^{01234}&=~2~T_{\mathrm{D4}}\\
    J_{\mathrm{D2}}^{012}~~~&=~J_{\mathrm{D2}}^{034}=0\\
    J_{\mathrm{D0}}^0~~~~&=~2~T_{\mathrm{D4}} (2\pi\alpha')^2 F_{12}F_{34}\,.
\end{align}
\end{subequations}
These brane currents do not preserve sixteen local supersymmetries; only eight supersymmetries are preserved, locally and globally. 

However, as we will demonstrate shortly, the currents \eqref{D0-D4_smeared} represent a singular limit where the local D2-brane densities on the two D4 branes exactly cancel each other. The cancellation occurs because the D2-brane densities are equal in magnitude (in order to preserve some global supersymmetry), but they point in opposite directions. By slightly moving onto the Coulomb branch (and separating the D4 branes), we will observe that there is a density of D2 and anti-D2 branes on each of the D4 branes. As such, the smeared instanton \eqref{D0-D4_smeared} can be interpreted as the juxtaposition of two abelian brane configurations, each possessing sixteen local supercharges, but with their dipolar charges perfectly cancelling each other. 

To prove this, let us turn on infinitesimally a scalar field, $\Phi$, which has the effect of slightly separating the two D4 branes along a spatial direction, $y$ , orthogonal to the worldvolume dimensions of the D4 branes. From the gauge-theory point of view, this corresponds to moving along the Coulomb branch. To have a non-trivial scalar field while preserving the same amount of supersymmetry, we need to set 
\begin{equation} 
    D_i \Phi =0 \,.
\end{equation}
One can require that $\Phi$ is constant; this leads to the constraint that $\Phi$ should commute with the gauge fields, and hence $\Phi = \zeta ~\sigma_3$.%
\footnote{Note that if we had turned on a scalar field $\Phi$ along $\Phi=\zeta~\sigma_{1}$ or $\zeta~\sigma_2$, not only this would not solve the equations of motions, but it would not preserve any supersymmetries. The intuitive reason is that with a scalar field different from the one along $\sigma_3$, the gauge fields that source our instanton would become massive. As a result, we would not see any local-supersymmetry structure even at finite $\zeta$.}
Then the density of branes in the momentum space dual to the position $y$ is given by:
\begin{subequations}
\begin{align}
    \tilde{j}_{\mathrm{D4}}^{01234}(k_y) &=~~~~~~~~~~T_{\mathrm{D4}} ~~\tr\left(e^{ik_y\Phi}\right)~~~~~~~~~=~T_{\mathrm{D4}}\left(e^{ik_y\zeta}+e^{-ik_y\zeta}\right)\\
    \tilde{j}_{\mathrm{D2}}^{012}(k_y) ~~&=~~~ T_{\mathrm{D4}} ~(2\pi\alpha') \tr\left(F_{34}\sigma_3 e^{ik_y\Phi}\right)~=~T_{\mathrm{D4}} (2\pi\alpha')F_{34}\left(e^{ik_y\zeta}-e^{-ik_y\zeta}\right)\\
    \tilde{j}_{\mathrm{D2}}^{034}(k_y)~~~ &= T_{\mathrm{D4}} ~(2\pi\alpha') \tr\left(F_{12}\sigma_3 e^{ik_y\Phi}\right)~~~~=~T_{\mathrm{D4}} (2\pi\alpha') F_{12}\left(e^{ik_y\zeta}-e^{-ik_y\zeta}\right)\\
    \tilde{j}_{\mathrm{D0}}^0(k_y) ~~&=T_{\mathrm{D4}} ~(2\pi\alpha')^2 \tr\left(F_{12}F_{34} e^{ik_y\Phi}\right)~=T_{\mathrm{D4}} (2\pi\alpha')^2 F_{12}F_{34}\left(e^{ik_y\zeta}+e^{-ik_y\zeta}\right)\,.
\end{align}
\end{subequations}
A trivial Fourier transform leads to:
\begin{subequations} \label{D4-D2-D0_density_coulomb}
\begin{align}
    j_{\mathrm{D4}}^{01234}(y)&=~T_{\mathrm{D4}} ~\left(\delta(y-\zeta)+ \delta(y+\zeta)\right) \\
    j_{\mathrm{D2}}^{012}(y)~~&=~ T_{\mathrm{D4}}~ (2\pi\alpha') F_{34} \left(\delta(y-\zeta)-\delta(y+\zeta)\right)\\
    j_{\mathrm{D2}}^{034}(y)~~&=~ T_{\mathrm{D4}} ~~(2\pi\alpha') F_{12} \left(\delta(y-\zeta)-\delta(y+\zeta)\right)\\
    j_{\mathrm{D0}}^0(y)~~&=~T_{\mathrm{D4}} ~(2\pi\alpha')^2 F_{12}F_{34}\left(\delta(y-\zeta)+ \delta(y+\zeta)\right)\,.
\end{align}
\end{subequations}
From this expression, we observe two copies of an abelian D4-D2-D2-D0 brane system (one on each branch $y=\pm \zeta$); however the D2-brane densities differ in their sign, indicating a different orientation. In the limit $\zeta \to 0$, we have $\delta(y - \zeta) - \delta(y + \zeta) = 0$, which results in the cancellation of the D2-brane densities. Thus, superposing the two branches together preserves only eight supersymmetries, locally and globally. In a nutshell, when the two D4 branes coincide, the dipolar charges exactly cancel each other, and thus there can no longer be sixteen local supersymmetries.

\subsection{A kappa-symmetry confirmation}
It has been shown in the context of a single brane that $\kappa$-symmetry can probe the brane densities \cite{Bena:2024oeq,Simon:2011rw}. This local fermionic symmetry of the full Born-Infeld action is not well understood in the non-abelian context, and only the first two orders in the field strength are known. Nevertheless we will see that $\kappa$-symmetry can probe the brane densities in a very efficient way, exactly reproducing the results of the previous section. 

In this instance, one has to be careful with the different Pauli matrices $\sigma$ that take part in the computation. One family of Pauli matrices are the $\mathrm{SU}(2)$ gauge-group generators: we will write them with a tilde (only the $\tilde{\sigma}_3$ is relevant for us). Another family of Pauli matrices appear in the construction of the brane involution matrix  (see Appendix \ref{appendix:projectors}). 
To fix notations, we choose, as in Section \ref{sec:Smeared instanton}, the smeared self-dual instanton:
\begin{equation}
    F=F_{12} \, \tilde{\sigma}_3 \, dx^1 \wedge dx^2 + F_{34}\,\tilde{\sigma}_3 \, dx^3 \wedge dx^4 \,.
\end{equation}

For such a field strength, we obtain from eq. \eqref{eq:kappa non abelian} the two $\kappa$-symmetry projection equations at quadratic order in the field strength (see eq. \eqref{eq:Gamma projection non commutative} of Appendix \ref{appendix:kappa_symm}) on the non-trivial component of the gauge-group valued Killing spinor $\varepsilon=\varepsilon_0 \mathbbm{1}$: 
\begin{subequations}
\begin{align}
    &\left[1+\sigma_3\,\Gamma^{01234}\left(\sigma_1+i\sigma_2\left(F_{12}\Gamma^{12}+F_{34}\Gamma^{34}\right)+\sigma_1(\Gamma^{1234}F_{12}F_{34}-\frac{1}{2}(F_{12}^2+F_{34}^2))\right)\right]\varepsilon_0=0\\
    &\left[1+\sigma_3\,\Gamma^{01234}\left(\sigma_1-i\sigma_2\left(F_{12}\Gamma^{12}+F_{34}\Gamma^{34}\right)+\sigma_1(\Gamma^{1234}F_{12}F_{34}-\frac{1}{2}(F_{12}^2+F_{34}^2))\right)\right]\varepsilon_0=0 \,.
\end{align}
\end{subequations}
Following the intuition from the abelian $\kappa$-symmetry\cite{Bergshoeff:2000ik}, a natural guess for the expression at full order in the field strength is:
\begin{subequations}
\begin{align}
    &\left[1+\frac{\sigma_3\,\Gamma^{01234}}{\sqrt{(1+F_{12}^2)(1+F_{34}^2)}}\left(\sigma_1+i\sigma_2\left(F_{12}\Gamma^{12}+F_{34}\Gamma^{34}\right)+\sigma_1\Gamma^{1234}F_{12}F_{34}\right)\right]\varepsilon_0=0\\
    &\left[1+\frac{\sigma_3\,\Gamma^{01234}}{\sqrt{(1+F_{12}^2)(1+F_{34}^2)}}\left(\sigma_1-i\sigma_2\left(F_{12}\Gamma^{12}+F_{34}\Gamma^{34}\right)+\sigma_1\Gamma^{1234}F_{12}F_{34}\right)\right]\varepsilon_0=0 \,.
\end{align}
\end{subequations}
Each of the two above equations describes the projection equation on a D4 brane carrying D2- and D0-brane charges. The D2-brane charges carried by the two D4 branes have opposite signs. This is consistent with the discussion in the previous subsection. Indeed, the equations match exactly the ones we obtained by slightly separating the branes \eqref{D4-D2-D0_density_coulomb}.

\subsection{The SU(2)-instanton structure revealed by kappa-symmetry}

In this subsection, we consider a general $\mathrm{SU}(2)$ instanton solution, a configuration which satisfies the self-duality condition \eqref{eq:self-duality}. And we will use the $\kappa$-symmetry projection equation to show why instantons do not have sixteen local supersymmetries, and highlight that generically the branes cannot even be separated while preserving supersymmetry. We conclude the subsection by explaining why we expect these instantons to be impossible to construct in supergravity.

Motivated by the intuition that $\kappa$-symmetry probes efficiently the branes densities, we compute order by order the projection operator. Following the procedure explained in Appendix \ref{appendix:kappa_symm}, we obtain to first order in the field strength:
\begin{equation}
    \Gamma_\kappa^{a 0\,(1)}=\Gamma^{(0)}\frac{1}{2} \Gamma^{ij}F_{ij}^{(a)}i \sigma_2 \,.
\end{equation}
Here, $a$ and $0$ in the superscript of $\Gamma_\kappa^{a 0\,(1)}$ label the generators of the Lie algebra and the identity respectively. $\Gamma^{(0)}$ is a product of Gamma matrices, defined below \eqref{eq:Gamma projection non commutative}.
To second order in the field strength, we have, after a more involved computation:
\begin{equation}
    \Gamma_\kappa^{a 0\,(2)}=\Gamma^{(0)}\sigma_1\delta^{a0}\left( \frac{1}{8}\Gamma^{ijkl}F_{ij}^{(c)}F_{kl}^{(c)}-\frac{1}{4} F_{kl}^{(c)}F^{kl\,(c)}\right)\,.
\end{equation}
We recall that $\sigma_1$ and $\sigma_2$ here are the Pauli matrices, but are not linked to the $\mathrm{SU}(2)$ gauge group, they are inherent to the construction of brane projection operators of Appendix \ref{appendix:projectors}. Recall that the supersymmetry generator is constant and Lie-algebra valued: $\varepsilon_a T^a$. Moreover, its only non-trivial component is along the identity (see Appendix \ref{appendix:kappa_symm}), we call $\varepsilon_0$ this component, and we obtain:
\begin{equation}
 \left( \mathbbm{1}+P^{a}\sigma_a
\right)\varepsilon_0 =0\,,
\end{equation}
where
\begin{equation}
{\footnotesize
      P^{a}\sigma_a \hspace{-0.3em}=\Gamma^{(0)}\hspace{-0.4em}\begin{bmatrix}
\sigma_1+\frac{i\sigma_2}{2}\Gamma^{ij}F_{ij}^{(3)}+ \frac{\sigma_1}{8}\Gamma^{ijkl}F_{ij}^{(c)}F_{kl}^{(c)}-\frac{1}{4} F_{kl}^{(c)}F^{kl\,(c)} & \frac{i\sigma_2}{2}\Gamma^{ij}\left(F_{ij}^{(1)}+i F_{ij}^{(2)} \right)\\
\frac{i\sigma_2}{2}\Gamma^{ij}\left(F_{ij}^{(1)}-i F_{ij}^{(2)} \right) & \sigma_1-\frac{i\sigma_2}{2}\Gamma^{ij}F_{ij}^{(3)}+ \frac{\sigma_1}{8}\Gamma^{ijkl}F_{ij}^{(c)}F_{kl}^{(c)}-\frac{1}{4} F_{kl}^{(c)}F^{kl\,(c)}
\end{bmatrix} . }
\label{bmatrix}
\end{equation}

There are several things to note in the above expression. First,  we specialise into the case in which all the spatial components of the gauge field are aligned along one of the gauge-group directions. This means that the field strength can be written as 
\begin{equation} \label{F_ij_aligned_gauge-group}
    F_{ij}^{(a)}~ T_a=F_{ij} ~T_0 \,,
\end{equation}
where $T_0$ is any element of the Lie algebra (normalized to $\tr(T_0^2)=1$), for all $(i,j)$. In this case, we can diagonalize \eqref{bmatrix}, so that it takes the form:
\begin{equation}
    P^{a}\sigma_a=\Gamma^{(0)}\begin{bmatrix}
\sigma_1+\frac{i\sigma_2}{2}\Gamma^{ij}F_{ij}+ \frac{\sigma_1}{8}\Gamma^{ijkl}F_{ij}F_{kl}-\frac{1}{4} F_{kl}F^{kl} & 0\\
0 & \sigma_1-\frac{i\sigma_2}{2}\Gamma^{ij}F_{ij}+ \frac{\sigma_1}{8}\Gamma^{ijkl}F_{ij}F_{kl}-\frac{1}{4} F_{kl}F^{kl}
\end{bmatrix} \,.
\end{equation}
This can be easily completed to all orders in the field strength by:
\begin{equation}
    P^{a}\sigma_a=\frac{1}{\sqrt{\det\left(\eta_{ij}+F_{ij}\right)}}\Gamma^{(0)}\begin{bmatrix}
\sigma_1+\frac{i\sigma_2}{2}\Gamma^{ij}F_{ij}+ \frac{\sigma_1}{8}\Gamma^{ijkl}F_{ij}F_{kl} & 0\\
0 & \sigma_1-\frac{i\sigma_2}{2}\Gamma^{ij}F_{ij}+ \frac{\sigma_1}{8}\Gamma^{ijkl}F_{ij}F_{kl}
\end{bmatrix} \,.
\end{equation}
In this case, we are back to a situation that is similar to the smeared instanton of Sec. \ref{sec:Smeared instanton}, with two projection equations. We thus have two branes that can be separated on their Coulomb branch. And each of the branes preserves independently sixteen local supersymmetries, with their respective dipolar charges cancelling each other when the two branes are bound.%
\footnote{In fact, while the smeared instanton was only a global analysis, here the equality $F_{ij}^{(a)} (x) \sigma_a=F_{ij}(x) T_0(x)$ can depend on the worldvolume coordinates, and therefore the diagonalization of \eqref{bmatrix} can be local.}

In general however, it is impossible to put all the spatial components of the field strength along the same gauge-group direction as we just did in \eqref{F_ij_aligned_gauge-group}. In this case, it is not even possible to separate the branes (\emph{i.e.} to turn on a constant scalar field $\Phi$) without breaking all the supersymmetries. Indeed in the general case, it is impossible to find a constant scalar field such that $D_i\Phi=0$.

In the original paper studying non-abelian $\kappa$-symmetry \cite{Bergshoeff:2000ik}, it was mentioned that it is impossible to factorize the projection equation by a $\mathrm{U}(2)$ singlet. We interpret this as the impossibility to split the branes without breaking supersymmetry. Indeed, in the cases where the gauge-group indices decouple from the field strength as in eq. \eqref{F_ij_aligned_gauge-group} (as for the smeared instanton), it was possible to factorize \eqref{bmatrix} by a $\mathrm{U}(2)$ singlet: $\det\left(\sqrt{\eta_{ij}+F_{ij}}\right)$. The impossibility to split the branes without breaking supersymmetry is linked to the impossibility to align all the field-strength spatial components along the same the gauge-group direction, and to go into the Coulomb branch for a generic instanton.

We have shown that instantons are by construction made of open strings  stretched between several branes in a complicated way. As a consequence, instantons come from non-abelian degrees of freedom, and by construction are non geometric. Unlike their monopole counterparts, they do not preserve sixteen local supersymmetries, and we interpret it as evidence that they cannot be constructed in supergravity as microstructure of smooth horizonless solutions. The two-charge D1-D5 solutions which have been constructed by Lunin and Mathur in supergravity \cite{Lunin:2001fv} have momentum and Kaluza-Klein momentum dipolar charges in the non-compact spatial dimensions, which make them geometric. By contrast, the gravity dual of instantons require D3-brane dipolar charges, and we do not expect these solutions to be constructible in supergravity.

\subsection{T-duality and link to monopoles}

We argued in Section \ref{sec:monopoles} that there is a well-defined notion of sixteen local supersymmetries for monopoles, but the previous subsections suggest that such a notion does not apply to instantons. This seems to raise an issue, since monopoles and instantons are related by T-duality (provided the assumption that one compactifies the extra dimension on a circle, and smears along this dimension). Indeed, starting with a (3+1)-dimensional monopole 
\begin{equation}
    (2\pi\alpha')\frac{\epsilon^{ijk}}{2}F_{jk}=D_i \Phi\,,
\end{equation}
one can perform a T-duality along the dimension, $x_4\equiv z$, that emerges out of the scalar field $\Phi$. Then, the resulting (4+1)-dimensional solution can be interpreted as an instanton solution with a spherical symmetry in $(x_1,x_2,x_3)$ and a translational invariance along $x_4$: the instanton solution takes the values of the monopole's field strength $F_{\mu\nu}$ on its (3+1) dimensions, and is supplemented by the identification 
\begin{equation}
    A_4(x_1,x_2,x_3,x_4) = \Phi(x_1,x_2,x_3) 
\end{equation}
on the fourth spatial dimension.

One could think that the difference between the T-dual monopole and instanton configurations is due to smearing, which is essential before T-dualising. Indeed, the need of smearing can be seen both from supergravity \cite{Bergshoeff:1995as} and from the abelian worldvolume action \cite{Bergshoeff:1996cy,Simon:1998az}. From our perspectives however, the reason is not due to smearing. Indeed, it is possible to perform T-duality on a circle without smearing from the non-abelian worldvolume action, by copying the branes living on a compact space an infinite number of times onto the covering space \cite{Taylor:1996ik}.

Similar to the computation of Section \ref{sec:monopoles}, we can compute the densities in momentum space by performing a non-abelian Fourier transform with respect to $A_4$ (which is identified with the scalar field $\Phi$):
\begin{equation}
    \tilde{j}_{\mathrm{D0}}^{0}=T_{\mathrm{D4}}(2\pi\alpha')^2\frac{\epsilon_{abc}}{2}\mathrm{STr}\left(e^{i (2\pi \alpha')k A_4} F_{ab}D_{c}A_4\right)\,.
\end{equation}
The above expression is exactly the one in eq. \eqref{eq:D1 brane density} for monopoles but with $A_4$ instead of $\Phi$.  Note that it is not an instanton on $\mathbb{R}^4$, but rather one on $\mathbb{R}^3\times \mathbb{S}^1$. Therefore, inserting the term $e^{i (\twopia) k A_4}$ in the trace (since $A_4$ is independent of $x_4$) is actually equivalent to adding a Wilson loop along the $x_4$ direction with $e^{i(\twopia)\frac{k}{2\pi}\int {\rm d}x_4 A_4}$. Then, by repeating the same computation as in the monopole section, we could think that we gave a local supersymmetry structure to the instanton: $A_4$ defined in such a way would allow us to understand the structure of the supersymmetries by identifying the unbroken $\mathrm{U}(1)^{N-1}$, and identify the Coulomb branch, like as we did for the monopoles. 
And it is indeed known that instantons on $\mathbb{R}^3\times \mathbb{S}^1$ are made of $N$ fundamental monopoles, and that is why a seemingly sixteen-local-supersymmetry structure would arise \cite{Lee:1998vu}.\footnote{The number of fundamental monopoles is $N$ rather than the expected $N-1$, because on a circle we can connect the $N^{\text{th}}$ brane to the $1^{\text{st}}$ one.} 

But there is a stark physical difference between $A_4$ in the instanton and $\Phi$ in the monopole, even though $\Phi$ is mapped to $A_4$ through T-duality.
The scalar field $\Phi\equiv \frac{1}{\twopia} X_4$ represents a physical direction, while $A_4$ is a gauge field. In the monopole case, the $e^{i(2\pi\alpha')k\Phi}$ in the trace in \eqref{D3D1densities_tHooftPolyakov} can be interpreted as a Fourier transform at an operator level, so that one can inverse-Fourier transform back with an abelian coordinate, $x_4$. However, in the instanton case, $A_4$ is not related to a space-time coordinate, so the na\"ive geometric interpretation is lost.

Moreover, in the monopole case, the $x_4$ direction is emergent, and on the emergent space, the dipolar charges live on different D3-brane branches and do not cancel. By contrast, the instanton solution $F_{ij}(x_1,x_2,x_3,x_4)$ has a translational invariance along $x_4$; therfore, one cannot use the $x_4$ direction to `separate' the dipole (D2-brane) charges.  As such, we conclude that one cannot reveal the sixteen-supersymmetries structure of instantons from a `bulk geometric' perspective.

Instantons on $\mathbb{R}^4$ are even more mysterious since we cannot turn on a Wilson line to identify a maximally abelian subgroup of the gauge group. It is in practice always possible to choose one gauge field, and repeat the procedure: non-abelian Fourier transform with respect to this gauge field, and go back into the real space in an abelian way. Any chosen gauge field would depend on all the world-volume coordinates: it is then even more unclear what the interpretation would be, since this gauge field cannot be interpreted as an extra dimension puffing up out of the brane.

Thus, it seems that the glues can only be revealed in a duality frame where the dipolar brane charges are separated in a physical space (as seen on the two D3 branches of the 't Hooft-Polyakov monopole). If this is not possible, then, the glue densities cancel and vanish, and there are no sixteen local supersymmetries. This is morally why some results are duality-frame dependent.

This duality dependence may have an important consequence. There are works that argue that the supergravity description of fuzzballs with internal excitations breaks down \cite{Kanitscheider:2007wq}, and collapses to the black-hole solution. The argument rests upon computing the supergravity solution with internal excitations through a chain of T- and S-dualities, starting from the F1-P frame. And as such, the solution depends on functions of \textit{one} variable. From our point of view however, one may not be able to reveal the dipolar properties of the internal excitation precisely because of T-duality, and this could be a reason why these supergravity solutions break down. We think one should determine the internal geometric structures directly in the relevant duality frame (\textit{e.g.} monopole frame) instead.

\vspace{1em}
All the smooth and horizonless geometries constructed so far in supergravity have sixteen local supersymmetries. As an example, the Lunin-Mathur geometries in the D1-D5 frame have sixteen local supersymmetries thanks to their dipolar KKM and momentum charges. Because the KK monopole takes a shape in the non-compact spatial dimensions, those dipolar charges geometrize the interaction between the D1 and D5 branes and make the system constructible in supergravity. But one can also write the D1-D5 system as instantons of the D5-brane worldvolume super Yang-Mills theory; the relevant dipoles would be D3-brane charges along the compact dimensions \cite{Eckardt:2023nmn}. However, we have just argued here that there is no natural way to reveal the D3-brane dipole charges and to detect the sixteen local supersymmetries in instanton configurations from a geometric perspective. This is because of the purely non-abelian degrees of freedom, and these cannot be described within supergravity.%
\footnote{There are supergravity solutions of localized D0 branes inside D4-branes \cite{Cherkis:2002ir}, but they are not dual to instantons. Indeed the D0 branes can be taken away from the D4 branes (Coulomb branch of the gauge theory) and hence the configuration does not form a bound state. Such solutions do not take into account the degrees of freedom of the gauge group of the higher-dimensional branes.}

By contrast, not only the D4-D2 system can also polarize into the same Kaluza-Klein-monopole and momentum dipolar charges as the D1-D5 system, but they can also polarise into D4-D2 dipolar charges wrapping the compact dimensions. The latter polarisation can be seen as monopoles of the super Yang-Mills theory on the D4-brane worldvolume. But this time, the non-abelian degrees of freedom of monopoles can be encoded from a geometric perspective.

Therefore, our viewpoint is that one should better construct microstates with internal excitations in duality frames where branes interact as monopoles (like the D2-D4-P and M2-M5-P frames) rather than in those where they interact as instantons (like the D1-D5-P frame).
In the following and last section, we show that the three-charge D2-D4-P building blocks have sixteen local supersymmetries, and reveal the dipole charges that are involved. This will be important data in light of the construction of their corresponding solutions in supergravity\cite{Bena:2024qed, Bena:2023fjx}.

\section{The three-charge monopole}
\label{sec:3-charge_monopole}

Type IIA string theory has been successful to count the entropy of the D2-D4-P black hole. A microscopic description at zero coupling is given by the fractionation of the D2 brane between the D4, each D2 strip become an independent momentum carrier\cite{Dijkgraaf:1996cv}. When including $\alpha'$ corrections (but still without turning on gravity), the D2 branes pull the D4 \cite{Hashimoto:1997px}, and are still independent momentum carrier. An analogous construction in M-theory for M2-M5-P black holes has shown that it should be possible to find such a bound state of branes with sixteen local supersymmetries \cite{Bena:2022wpl}. The 't Hooft-Polyakov monopole can be interpreted as being such a D2/M2 strip, on which we would like to add momentum. In Section \ref{sec:monopoles}, we have shown that all monopoles of Yang-Mills theory have sixteen local supercharges. 

Monopoles in Section \ref{sec:monopoles} are (3+1) dimensional, on the world-volume theory of D3 branes; they describe the physics of orthogonal D3-D1 branes when they start interacting. One can `add' an extra dimension and consider the equivalent D4-D2 solution, with nothing special happening in the extra dimension. The D1 become D2 strips with a common direction with the D4 branes. Our purpose is to add momentum on this extra dimension. It has been shown in \cite{Bena:2023fjx} that such a construction is possible and preserves sixteen local supersymmetries in its abelian version, and has been fully backreacted in \cite{Bena:2024qed}. It has been shown that we can add a momentum wave to the 't Hooft-Polyakov monopole without changing its shape\cite{Bena:2023fjx}, we first review this construction, then we compute the branes densities and show that this three-charge monopole preserves sixteen local supersymmetries.

\subsection{The three-charge solution}

One can superpose a travelling momentum wave onto the ’t Hooft–Polyakov monopole without deforming it\cite{Bena:2023fjx}. The 't Hooft-Polyakov monopole is a solution of $\mathcal{N}=4$ SYM in 3+1 dimensions, which describes a D1 brane suspended between two D3s. In (4+1)-dimensional maximally supersymmetric Yang-Mills, it is still a solution, with the fourth spatial direction untouched. It describes a D2 brane strip suspended between two D4 branes, and the different branes share the fourth spatial dimension that we call $y$. We can add momentum on this direction without changing the shape of the monopole. This results in the computation done for D1 strip between two D3 being exactly the same but where the tension is now that of a D4 brane, and the densities of D1 (resp. D3) become the densities of D2 (resp. D4) by the addition of the extra direction $y$. 

Let $f(y-t)$ a periodic function which averages to zero. The momentum is carried by turning on two non-trivial components of the gauge field
\begin{align}
	A_{0}^{a}=\Phi^{a}f(y-t)\,, \quad A_{y}^{a}=-\Phi^{a}f(y-t)\,,
\end{align}
on top of the original 't Hooft-Polyakov monopole solution of Section \ref{sec:monopoles}:
\begin{align}
	&A_{i}^{a}=\epsilon_{aij}\frac{x^{j}}{r}\frac{(1-K(r))}{r} \\
	&\Phi^{a}=\frac{x^{a}}{r^2}H(r) \,.  
\end{align}
This results in keeping the field strength $F_{\mu\nu}$ describing the shape of the (3+1)d monopole untouched, to which one adds two new non-trivial field-strength components:
\begin{equation} \label{new_components_F2}
    F_{yr}=F_{r0}\,.
\end{equation}

\begin{figure}[h!]
    \centering
    \includegraphics[scale=1]{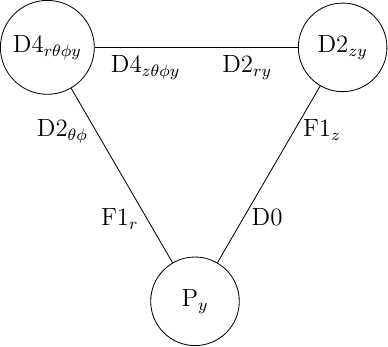}
    \caption{Triality of the brane densities. The global brane charges are at the corners of the triangle. Each pair of global brane charges is linked by a pair of glues.} 
    \label{fig:D4-D2-P_diagram}
\end{figure}

Before computing all the charges, it is useful to summarize the list of expressions to which the relevant objects in string theory couple to (Cf. Section \ref{ssec:brane charges single stack}):%
\footnote{As explained in Section \ref{sec:brane density formalism}, the densities of momentum and fundamental strings are not expected to be the most general expressions, but rather their linearized form in $\alpha'$. Nevertheless, for BPS objects such as our three-charge bound state, the linearized and the full expression are expected to be the same.}
\begin{equation} 
    \begin{matrix}
        &[C_{0 r\theta\phi y}] \, 1&\rightarrow & \mathrm{D4}\, ( r\theta\phi y) \,,& 
        &[C_{0 z\theta\phi y}] \, \partial_r \Phi &\rightarrow& \mathrm{D4}\, ( z\theta\phi y)\,,\\
        &[C_{0 r y}]\,F_{\theta\phi}&\rightarrow& \mathrm{D2}\, ( r y) \,,& 
        &[C_{0zy}]\,F_{\theta\phi}\partial_r \Phi& \rightarrow& \mathrm{D2}\, ( z y)\,,\\
        &[C_{0\theta\phi}]\,F_{yr}&\rightarrow &\mathrm{D2}\, (\theta\phi) \,,& 
        &F_{0r}&\rightarrow& \mathrm{F1} \,(r) \,,\\
        &F_{0r}\partial_r \Phi&\rightarrow& \mathrm{F1} \,(z) \,, & 
        & [C_{0}]\, F_{ry}F_{\theta\phi} &\rightarrow &\, \mathrm{D0} \,,\\
        &F_{0r}F_{ry}&\rightarrow& \mathrm{P}\,(y) \,.
    \end{matrix}
\end{equation}
In fact, three of the nine brane densities are that of the `global brane charges' of the three-charge D4-D2-P system, while six of them are densities of `glues' (dipoles). As we will later show, the values of densities of global and dipolar charges are constrained, and it is convenient to represent them in an organised form in Fig. \ref{fig:D4-D2-P_diagram}.

The top part of the diagram in Fig. \ref{fig:D4-D2-P_diagram} is simply Fig. \ref{fig:D3-D1_diagram} in Type IIA theory, and the densities of the glues are constrained to be equal through the monopole equation \eqref{BPS-eq_tHP}. On the bottom left and bottom right parts of the diagram, the glue densities are locked thanks to equations \eqref{BPS-eq_tHP} and \eqref{new_components_F2}.

\subsection{The brane densities}

We can now compute the densities of momentum, fundamental strings, and branes. We choose to compute the densities from the bottom to the top of Fig. \ref{fig:D4-D2-P_diagram}. 

First, the density of momentum along $y$ is computed through the formula
\begin{equation}
\tilde{j}_{\mathrm{P}}^{0y}(k_z,r,y-t) =T_{\mathrm{D4}}(2\pi\alpha')^2\mathrm{STr}\left(e^{i (2\pi \alpha')k_z\Phi} F_{0i}^{a}{F^{a}}^{iy}\right)\,,
\end{equation}
which yields, in coordinate space, to
\begin{align}
    j_{\mathrm{P}}^{0y}(z,r,y-t)=~&  T_{\mathrm{D4}}(2\pi\alpha')^2 f(y-t)^2\biggl[\frac{(1-K^2)^2}{4 r^4}\left( \delta\left(z-\lambda(r)\right)+\delta\left(z+\lambda(r)\right)\right) \nonumber\\
    &\hspace{9em}+\frac{H K^2}{2\pi \alpha' r^3}\left( \Theta\left(z-\lambda(r)\right)-\Theta\left(z+\lambda(r)\right)\right)\biggr] \,.
\end{align}

Then, the densities of the glues, respectively on the left and right edges of the diagram in Fig. \ref{fig:D4-D2-P_diagram}, are given by the expressions:
\begin{subequations}
\begin{align}
    &\tilde{j}_{\mathrm{F1}}^{0r} =T_{\mathrm{D4}}(2\pi\alpha')\frac{x_{i}}{r}\mathrm{STr}\left(e^{i (2\pi \alpha')k_z \Phi} F_{0i}\right)\\
    &\tilde{j}_{\mathrm{D2}}^{0\theta\phi} = T_{\mathrm{D4}}(2\pi\alpha')\frac{x_{i}}{r}\mathrm{STr}\left(e^{i (2\pi \alpha')k_z\Phi} F_{iy}\right)
\end{align}
\end{subequations}
\begin{subequations}
\begin{align}
    &\tilde{j}_{\rm F1}^{0z} =T_{\mathrm{D4}}(2\pi\alpha')^2\mathrm{STr}\left(e^{i (2\pi \alpha')k_z \Phi} F_{0i}D_{i}\Phi\right)\\
    &\tilde{j}_{\mathrm{D0}}^{0} =T_{\mathrm{D4}}(2\pi\alpha')^2\frac{\epsilon_{ijk}}{2}\mathrm{STr}\left(e^{i (2\pi \alpha')k_z \Phi} F_{ij}F_{ky}\right)\,.
\end{align}
\end{subequations}
We compute the symmetrised traces, and obtain, back in position space:
\begin{subequations}
\begin{align}
    j_{\mathrm{D2}}^{0\theta \phi}=~& T_{\mathrm{D4}}(2\pi\alpha') f(y-t)\frac{1-K^2}{2 r^2} \bigl[ \delta\left(z-\lambda(r)\right)-\delta\left(z+\lambda(r)\right)\bigr]\\
    j_{\mathrm{F1}}^{0r}=~& T_{\mathrm{D4}}(2\pi\alpha')f(y-t)\frac{1-K^2}{2 r^2} \bigl[ \delta\left(z-\lambda(r)\right)-\delta\left(z+\lambda(r)\right)\bigr]
\end{align}
\end{subequations}
\begin{subequations}
\begin{align}
    j_{\mathrm{F1}}^{0z}=~ & T_{\mathrm{D4}}(2\pi\alpha')^2 f(y-t)\biggl[\frac{(1-K^2)^2}{4 r^4}\left( \delta\left(z-\lambda(r)\right)+\delta\left(z+\lambda(r)\right)\right) \nonumber\\
    &\hspace{9em} +\frac{H K^2}{2\pi \alpha' r^3}\left( \Theta\left(z-\lambda(r)\right)-\Theta\left(z+\lambda(r)\right)\right)\biggr]\\
    j_{\mathrm{D0}}^{0}=~ & T_{\mathrm{D4}}(2\pi\alpha')^2 f(y-t)\biggl[\frac{(1-K^2)^2}{4 r^4}\left( \delta\left(z-\lambda(r)\right)+\delta\left(z+\lambda(r)\right)\right) \nonumber\\
    &\hspace{9em} +\frac{H K^2}{2\pi \alpha' r^3}\left( \Theta\left(z-\lambda(r)\right)-\Theta\left(z+\lambda(r)\right)\right)\biggr]\,.
\end{align}
\end{subequations}

Finally, the four brane densities of the top of the diagram of Fig. \ref{fig:D4-D2-P_diagram} have been computed in eq. \eqref{D3-D1-densities_position_tHP}:
\begin{subequations} 
\begin{align}
    j_{\mathrm{D4}}^{0r\theta\phi}=~& T_{\mathrm{D4}} \left[ \delta\left(z-\lambda(r)\right)+\delta\left(z+\lambda(r)\right)\right]\\
    j_{\mathrm{D4}}^{0z\theta\phi}=~& T_{\mathrm{D4}}(2\pi\alpha') \frac{1-K^2}{2 r^2} \left[ \delta\left(z-\lambda(r)\right)-\delta\left(z+\lambda(r)\right)\right] \\
    j_{\mathrm{D2}}^{0r}=~& T_{\mathrm{D4}}(2\pi\alpha')\frac{1-K^2}{2 r^2} \left[ \delta\left(z-\lambda(r)\right)-\delta\left(z+\lambda(r)\right)\right]\\
    j_{\mathrm{D2}}^{0z}=~ & T_{\mathrm{D4}}(2\pi\alpha')^2 \biggl[\frac{(1-K^2)^2}{4 r^4}\left( \delta\left(z-\lambda(r)\right)+\delta\left(z+\lambda(r)\right)\right) \nonumber\\
    &\hspace{7em} +\frac{H K^2}{2\pi \alpha' r^3}\left( \Theta\left(z-\lambda(r)\right)-\Theta\left(z+\lambda(r)\right)\right)\biggr] \,.
\end{align}
\end{subequations}

Just as the momentum-less 't Hooft-Polyakov monopole, the currents can be spatially divided into three different branches:
\begin{itemize}
    \item The upper branch on the D4 location, parametrized by $\delta\left(z-\lambda(r)\right)$. The nine brane densities (three main branes, 6 glues) are present on this branch. 
    \item The lower branch on the D4 location, parametrized by $\delta\left(z+\lambda(r)\right)$. The same nine brane densities are present, but some of them have a sign difference with their upper-branch counterparts. 
    \item The region in between the two D4 branes, characterized by $\Theta\left(z-\lambda\right)-\Theta\left(z+\lambda\right)$. There are only four brane densities here, which, as we will shortly see, behave like a two-charge solution of a D2 brane carrying momentum.
\end{itemize}

\subsection{The supersymmetries on top of the D4 branes}

We consider the brane densities on the upper branch of the solution, and extract the coefficient in front of $\delta\left(z-\lambda(r)\right)$ in the computations. As in Section \ref{ssec:tHooft-Polyakov}, we define the function $\chi$, corresponding to the tilt of the D4 brane on the upper branch:
\begin{equation}
\chi(r) \equiv (2\pi\alpha')\frac{1-K(r)^2}{2 r^2} = {\partial_rz \,}_{|\mathrm{upper \, branch}} = +\frac{\mathrm{d}\lambda}{\mathrm{d}r} \,.
\end{equation}

Thus, the densities of the main branes on the upper branch simplify to
\begin{align} \label{density_three_main_branes}
j_{\mathrm{D4}}^{0r\theta\phi y}= T_{\mathrm{D4}} \,, \qquad 
j_{\mathrm{D2}}^{0zy}= T_{\mathrm{D4}} \chi(r)^2 \,, \qquad 
j_{\mathrm{P}}^{0y}=T_{\mathrm{D4}} \chi(r)^2 f(y-t)^2 \,,
\end{align}
and the densities of the glues to
\begin{subequations}
\begin{align}
j_{\mathrm{D4}}^{0z\theta\phi y} = j_{\mathrm{D2}}^{0ry}&=T_{\mathrm{D4}}\chi(r) \\
j_{\mathrm{D2}}^{0\theta\phi}= j_{\mathrm{F1}}^{0r}&= T_{\mathrm{D4}} \chi(r) f(y-t) \\
j_{\mathrm{F1}}^{0z}= j_{\mathrm{D0}}^{0} &=T_{\mathrm{D4}} \chi(r)^2 f(y-t) \,.
\end{align}
\end{subequations}
Note that the density of each pair of glues is, up to a sign, the geometric mean of the densities of its adjacent main branes in Fig. \ref{fig:D4-D2-P_diagram}. 
On the lower branch, the densities of the main branes are the same, while the densities of the glues become
\begin{subequations}
\begin{align}
j_{\mathrm{D4}}^{0z\theta\phi y} = j_{\mathrm{D2}}^{0ry}&= -T_{\mathrm{D4}}\chi(r) \\
j_{\mathrm{D2}}^{0\theta\phi}= j_{\mathrm{F1}}^{0r}&= -T_{\mathrm{D4}} \chi(r) f(y-t) \\
j_{\mathrm{F1}}^{0z}= j_{\mathrm{D0}}^{0} &= +T_{\mathrm{D4}} \chi(r)^2 f(y-t) \,.
\end{align}
\end{subequations}
Therefore, the dipolar nature of the glues has two origins. The first, shared by the two former pairs of glues, is that the upper and lower branches have opposite brane densities. The second, shared by the two latter pairs of glues, comes from the fact that $f(y-t)$ is sometimes positive, sometimes negative, but averages to zero. 

Importantly, the multipolar nature of the brane densities illustrates the difference between the black-hole solution and the black-hole microstates. What we have described above is a simple model of a black-hole microstate. When one integrates the densities over the compact spatial direction, the only non-vanishing brane charges are the three charges of the five-dimensional black-hole, parametrised by $(N_2,N_4,N_{\rm P})$. (For example, $N_{\rm P}=\int \mathrm{d}^5x \, j_{\mathrm{P}}^{0y}$.) Therefore, the black-hole solution only knows about the average value of the brane densities, while the microstates possess a rich multipolar structure.

As in Section \ref{ssec:tHooft-Polyakov}, one can compute the Hamiltonian density on either branch:
\begin{equation} 
    \mathcal{H}_{\mathrm{ontop}}(r,y-t) =T_{\mathrm{D4}} \left(1 +\chi(r)^2 + \chi(r)^2 f(y-t)^2\right)\,.
\end{equation}
Thus, one recognises that the densities of the main branes \eqref{density_three_main_branes} relative to the Hamiltonian density describe a $\mathbb{S}^2$ of unit radius. Putting the nine brane densities $j_i$ together, it is not difficult to show that $\Pi(r,t-y)\equiv \frac{1}{2}\left(1+\sum_i \frac{j_i}{\mathcal{H}_{\rm ontop}} P_i\right)$ (where the $P_i$ are the involutions as in \eqref{projector_local_currents}) is a projector \cite{Bena:2022wpl}. This shows that everywhere on the D4-brane locus, the brane densities exhibit sixteen local supersymmetries.

To determine the global charges it is convenient to parametrise the $\mathbb{S}^2$ with the angles $\alpha,\beta$ defined as follows \cite{Bena:2023fjx,Bena:2022wpl}:
\begin{equation}
    \tan\alpha = \partial_rz = \pm \chi(r) \,,\quad 
    \tan\beta=\frac{(\partial_rz) f(y-t)}{\sqrt{1+\chi(r)^2}} = \pm\frac{\chi(r) f(y-t)}{\sqrt{1+\chi(r)^2}} \,, 
\end{equation}
and one can factorise the projection equation in the following fashion, like \eqref{split_projector_mainbranes}:
\begin{align}
    &\cos\alpha\left(\cos\alpha\cos^2\beta+\sin\alpha\cos^2\beta \,\Gamma^{zr}+\cos\beta\sin\beta\,\Gamma^{ry}\sigma_3\right)\left(1+\Gamma^{0r\theta\phi y}i\sigma_2\right)\varepsilon \nonumber\\ 
	&+\sin\alpha \left(\sin\alpha\cos^2\beta-\cos\alpha\cos^2\beta\,\Gamma^{zr}+\cos\beta\sin\beta\,\Gamma^{zy}\sigma_3)(1+\Gamma^{0zy}\sigma_1\right)\varepsilon \nonumber\\ 
	&+ \sin\beta\left(\sin\beta-\cos\alpha\cos\beta\,\Gamma^{ry}\sigma_3-\sin\alpha\cos\beta\,\Gamma^{zy}\sigma_3\right)(\Gamma^{0y}+1)\varepsilon=0\,.
\end{align}
This yields three projection equations that are independent of the space-time coordinates, namely the projection equations of the three main branes.

\subsection{The supersymmetries between the branes}

\begin{figure}[h!]
    \centering
    \includegraphics[scale=1]{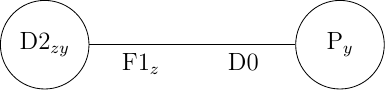}
    \caption{Brane densities in between the D4 branes.}
    \label{fig:D2-P_diagram}
\end{figure}

For the densities between the two branches, it is convenient to define
\begin{equation}
    \xi(r)=(2\pi\alpha') \frac{H K^2}{r^3} \,.
\end{equation}
Then the brane densities are given by:
\begin{subequations}
\begin{align}
j_{\mathrm{D2}}^{0zy}(z,r,t-y)&= T_{\mathrm{D4}} \xi(r) \,, \qquad 
j_{\mathrm{P}}^{0y}(z,r,t-y)=T_{\mathrm{D4}} \xi(r) f(y-t)^2 \,,\\
&\hspace{-3em} j_{\mathrm{F1}}^{0z}(z,r,t-y)=j_{\mathrm{D0}}^{0}(z,r,t-y)=T_{\mathrm{D4}} \xi(r) f(y-t) \,.
\end{align}
\end{subequations}
And here the density of energy is given by:%
\footnote{The total energy density in between the two D4 branes is $\mathcal{H}_{\mathrm{btw}}(r,t-y)=\int \mathrm{d}z\,\tilde{\mathcal{H}}_{\mathrm{btw}}(z,r,t-y)= 2 T_{\mathrm{D4}}\xi(r) \frac{H(r)}{r} \left(1+f(y-t)^2\right) $.}
\begin{equation}
    \tilde{\mathcal{H}}_{\mathrm{btw}}(z,r,t-y)= T_{\mathrm{D4}}\xi(r) \left(1+f(y-t)^2\right)\,.
\end{equation}
Again, the energy density is the sum of the densities of the two main branes. The brane densities present in the region between the D4 branes follow a main-brane/glue structure summarised in Fig. \ref{fig:D2-P_diagram}.
They correspond to the lower-right part of the three-charge diagram in Fig. \ref{fig:D4-D2-P_diagram}.

The projection equation in the region between the D4 branes is
\begin{equation}
    \left[1+\frac{1}{1+f(y-t)^2}\left(\Gamma^{0zy}\sigma_1+f(y-t)\,\Gamma^{0}i\sigma_2 -f(y-t)\,\Gamma^{0z}\sigma_3 +f(y-t)^2\,\Gamma^{0y}\right)\right]\varepsilon=0\,.
\end{equation}
This is the projection equation a D2 brane carrying momentum with longitudinal D0 and F1 excitation, and admits sixteen local supersymmetries. Note that all dependence in the radius $r$ factors out in this equation.

\vspace{1em}
In this section, we showed that one can add momentum to monopoles, and the number of local supersymmetries is still sixteen. The property of sixteen local supersymmetries suggests that supergravity solutions matching the geometry of monopoles with a momentum can be constructed. These solutions should correspond to the backreaction (in the gravity regime) of the `supermaze' solutions \cite{Bena:2022wpl}, that are microstates of a three-charge (D4-D2-P) black hole. From the gauge-theory perspective, we have computed the brane densities involved in a D4-D2-P microstate. We have also shown that the brane densities are split between a part that is bound to the worldvolume of the D4 branes, and a part that lies in between the branes. Interestingly, the supersymmetry structure bound to the D4-brane worldvolume is the one represented in the diagram of Fig. \ref{fig:D4-D2-P_diagram}, while that of the region in between the branes is the one in the diagram of Fig. \ref{fig:D2-P_diagram}. These interesting features constitutes essential data for the construction of supersymmetric black-hole microstates and for the Fuzzball programme.

\section{Conclusion and outlook}
\label{sec:conclusion}

In this article, we have studied the local supersymmetries of a stack of branes, from their low-energy effective action, namely the Dirac-Born-Infeld action. To this effect, we have computed, in various examples, the brane densities that are encoded in the SYM theory of the stack of branes. We studied situations where D-branes interact as to generate monopole solutions (Section \ref{sec:monopoles}), instanton solutions (Section \ref{sec:instantons}), and solutions involving higher-dimensional branes (Section \ref{sec:brane_polarization}). We proved the existence of sixteen local supersymmetries for monopole solutions and for solutions involving the polarisation into higher-dimensional branes, and we have demonstrated that instantons lack this property. We have also shown that it is possible to add a momentum wave to monopole solutions without disrupting their local supersymmetry structure (Section \ref{sec:3-charge_monopole}); this suggests that they may be constructible in supergravity.

The novel aspect of our article is the study of the non-abelian degrees of freedom carried by a stack of branes. We have demonstrated that BPS solutions exhibit an underlying local supersymmetry structure when the gauge group is maximally broken by a scalar field. In the examples with monopoles and brane polarisations, the solutions can be interpreted as a superposition of abelian branches. This is the reason why we think these degrees of freedom may be captured by supergravity. Conversely, instantons can rarely be split into abelian branches, so they seem to be pure stringy objects, inaccessible from supergravity.

This study is crucial in the understanding of the geometric structures of BPS black-hole microstates. Indeed, the brane densities we compute involve dipolar currents that are characteristic of pure black-hole microstates. The black-hole solution, on the other hand, has only information about the averaged values of the brane densities. Besides, the existence of sixteen local supersymmetries is a smoking gun for the existence of solutions similar to black holes, but without an event horizon \cite{Bena:2022wpl,Bena:2011uw,Bena:2022fzf}. Thus, the difference of the local supersymmetry structure between monopoles (D2-D4-P systems) and instantons (D1-D5-P systems) suggests that \textit{the ability to find horizonless microstates of BPS black holes in string theory depends on the duality frame!}

Because we have shown that instantons in the D5-brane action \cite{Maldacena:1996ky} cannot preserve sixteen local supercharges, we conjecture that it is impossible to construct D1-D5 geometries in supergravity with dipolar D3 branes.
On the other hand, the properties of the brane densities we compute in Section \ref{sec:3-charge_monopole} give us crucial information to construct supergravity solutions of microstates of a D2-D4-P black hole, with a finite horizon area. The waves on the monopoles we study carry only internal excitations, with no extension in the non-compact spatial dimensions, so these solutions should be qualitatively different from all Fuzzball solutions constructed to date. (Indeed,  all microstate geometries constructed from supergravity rely on the presence of a Kaluza-Klein-monopole dipole, whose shape in the external dimensions makes the solution geometric\cite{Lunin:2001fv,Bena:2015bea}.)  

For future work, we would like first to test whether momentum fractionation can be compatible with the local supersymmetry structure of the three-charge monopoles of Section \ref{sec:3-charge_monopole}. The idea is to make the monopole configuration vary as one moves along the common D2-D4 circle. The periodic boundary conditions on the circle impose a mapping between the D2 strips, and generate momentum fractionation. 
The second question is about the backreaction in the supergravity regime of such monopoles with momentum. We expect monopole configurations to be characterised by topological numbers that tell how the D2 branes are split inside the D4 branes. The idea is to track these numbers into the supergravity regime through geometric transition. 
We intend to explore these directions to investigate black-hole microstates at finite gravitational coupling.

\vskip 20pt
	\noindent {\bf Acknowledgements:} We would like to thank Iosif Bena, Antoine Bourget, Nejc Ceplak, Emil Martinec, Luca Martucci, Masaki Shigemori and Wati Taylor for interesting discussions. 
 The work of YL is supported by the University of Padua under the 2023 STARS Grants@Unipd programme (GENSYMSTR – Generalized Symmetries from Strings and Branes) and in part by the Italian MUR Departments of Excellence grant 2023-2027 "Quantum Frontiers”.
\clearpage

\appendix
\section{List of traceless involutions for branes}
\label{appendix:projectors}

The traceless involutions matching the different types of branes are:
\begin{equation}\nonumber
	\begin{array}{l|l}
		P_{\rm P}=\Gamma^{01} & P_{\rm  F1}=\Gamma^{01}\sigma_3\\ 
		P_{\rm D0}=\Gamma^{0}i\sigma_2 & P_{\rm D1}=\Gamma^{01}\sigma_1\\
		P_{\rm D2} =\Gamma^{012}\sigma_1 & P_{\rm D3}=\Gamma^{0123}i\sigma_2\\
		P_{\rm D4}=\Gamma^{01234}i\sigma_2 &P_{\rm  D5}=\Gamma^{012345}\sigma_1\\
		P_{\rm D6}=\Gamma^{0123456}\sigma_1&P_{\rm D7}=\Gamma^{01234567}i\sigma_2 \,.
	\end{array}
\end{equation}

\section{Kappa-symmetry}
\label{appendix:kappa_symm}

In this Appendix, following Section \ref{sec:kappasym Non abelian}, we write up the technical details of $\kappa$-symmetry needed for the article.

The precise definition of the non-abelian version of the Born-Infeld action is poorly understood, mainly because of the difficulty to differentiate between higher-derivative corrections and $\alpha'$ corrections. Indeed, both types of corrections are mixed up through the identity $[D_i,D_j]F_{kl}=-\frac{i}{2}[F_{ij},F_{kl}]$ \cite{Tseytlin:1997csa}. As a consequence, the non-abelian version of $\kappa$-symmetry is only known up to second order in $F_{ij}$. However, as we will see, it is still a powerful tool to probe brane densities. 

First, let us describe how to compute the Killing-spinor equation up to second order in the field strength, following \cite{Bergshoeff:2000ik,Bergshoeff:2001dc}. We will review this with a $\mathrm{SU}(2)$ gauge group in the irreducible representation of dimension $N$. Tseytlin proposed in \cite{Tseytlin:1997csa} a generalisation of the Born-Infeld action for non-abelian configurations, based on a symmetrized-trace prescription to solve an ambiguity in the ordering of the terms:
\begin{equation}
    S_\mathrm{BI} =-T_{\mathrm{D}p} \int d^{p+1}\sigma \, \str\sqrt{-{\rm det}\left(\eta_{\mu\nu}+ 2\pi \alpha' \,F_{\mu\nu} \right)}\,.
\end{equation}
This action reduces to the super-Yang-Mills theory at first $\alpha'$ order.
The total low-energy action for the brane has to be supplemented with the topological Wess-Zumino term \eqref{non-commutative_WZ} which governs the coupling to other kind of branes.

The total action is expected to have a $\kappa$-symmetry, resulting in a projection equation for the supersymmetry parameter with at most sixteen non-trivial solutions. $\kappa$-symmetry then removes half of the fermionic degrees of freedom . In the abelian case, we can define a notion of local supersymmetry: by zooming at a precise location on the brane, we could in principle T-dualise the solution to a single stack of brane \cite{Bena:2022wpl,Bena:2022fzf}.

In the case of the abelian Born-Infeld action, $\kappa$-symmetry results in a projection equation for a constant supersymmetry parameter $\varepsilon$ of the form:
\begin{equation}
    \left(1-\Gamma_{\kappa}\right)\varepsilon=0 \,.
\end{equation}
In the non-abelian Born-Infeld action, we have the same expression but with both the supersymmetry parameter $\varepsilon=\varepsilon_{a}T^{a}$ (where $T^{a}$ are the generator of the Lie algebra) and the $\Gamma_\kappa$ matrix being gauge-group valued. The projection equation then takes the form
\begin{equation}
\label{eq:kappa non abelian}
    \left(\mathbbm{1} \delta^{ab}-\Gamma^{ab}_{\kappa}\right)\varepsilon_{b}=0\,,
\end{equation}
with $a,b$ being the gauge-group indices. Besides, the projection matrices must satisfy:
\begin{equation}
    \Gamma_\kappa^{ab}\Gamma_\kappa^{bc}=\delta^{ac}\,.
\end{equation}

The supersymmetry generator $\varepsilon$ is constant, and when we commute it with a Yang-Mills gauge transformation we obtain:
\begin{equation}
    f^{abc}\varepsilon_{c}=0\,,
\end{equation} 
with $f^{abc}$ being the structure constant of the gauge group. Thanks to this condition, we deduce that $\varepsilon$ is along the identity: The only non-trivial supersymmetry parameter is thus $\varepsilon=\varepsilon_0 \mathbbm{1}$.

The projection matrix was computed order by order up to second order in the field strength and is given by:
\begin{align} \label{eq:Gamma projection non commutative}
    \Gamma_\kappa^{ab}=\Gamma^{(0)}\biggl\{&\delta^{ab}\sigma_1+P_{(1)}^{abc}\frac{1}{2}\gamma^{ij}F_{ij}^{(c)}\\
    &-\sigma_{1} \mathcal{S}^{abcd}\left(\frac{1}{8}\gamma^{ijkl}F_{ij}^{(c)}F_{kl}^{(d)}-\frac{1}{4}F_{kl}^{(c)} F^{kl\, (d)} \right) \nn\\
    &-\sigma_1 \mathcal{A}^{abcd}\gamma^{ij}F_{ik}^{(c)}F_j^{k \, (d)}\biggr\} \,. \nn
\end{align}
where the $\gamma^{ij}$ are the previous $\Gamma^{ij}$ gamma matrices (here we write them with a lower-case letter to avoid confusion with $\Gamma^{ab}_{\kappa}$) and where $\Gamma^{(0)}=\frac{1}{(p+1)!} \epsilon^{i_0 \dots i_p}\gamma_{i_0\dots i_p}$.
First, we recall that $f_{abc}$ is the structure constant of the Lie algebra. Moreover, the remaining quantities of the above equation are defined out of the commutation relations of the generators $T^a$:
\begin{align}
    &\{T_a,T_b\}=2 \, d_{abc} \,T_c\\
    &P_{(1)}^{abc}=i\sigma_2\, d^{abc}\\
    &\mathcal{S}^{abcd}=-d^{ae(c}d^{bd)e}\\
    &\mathcal{A}^{abcd}=-d^{ae[c}d^{bd]e} \,,
\end{align}
where symmetrisation and anti-symmetrisation in the last two lines are only performed over $c$ and $d$. We use conventions such that:
\begin{equation}
    \tr (T_{a}T_{b})=\delta_{ab} \,.
\end{equation}
Note that those conventions are different from the normalisation of the generators of the gauge groups in the bulk of the text. This difference does not matter at the level of the projection equations. It has been shown that the $\kappa$-symmetry projection equation \eqref{eq:kappa non abelian} cannot be factorized by a $\mathrm{U}(N)$ singlet (such as the on-shell Lagrangian), as was possible in the abelian case\cite{Bergshoeff:2000ik}. In Section \ref{sec:instantons}, we shed light on this phenomenon and provide a physical interpretation in terms of local supersymmetry.
We choose the gauge group to be $\mathrm{U}(2)$. One obtains, for the irreducible representation of $\mathrm{U}(2)$ of dimension $N$, the relation:
\begin{equation}
    d^{abc}=\frac{1}{\sqrt{N}}\left(\delta^{ab}\delta^{c0}+\delta^{cb}\delta^{a0}+\delta^{ac}\delta^{b0}-2\delta^{0a}\delta^{0b}\delta^{0c}\right)\,.
\end{equation}

\section{Brane densities for the most general monopole}
\label{sec:appendix:brane densities general monopole}
In this appendix we compute the D1-brane density along $z$ for the most general monopole, see Section \ref{ssec:monopoles_general}. The D1-brane density in momentum space is given by:
\begin{equation}
    j_{\mathrm{D1}}^{0z}(k_z)=T_{\mathrm{D3}}(2\pi \alpha')^2 \frac{\epsilon^{ijk} }{2}\mathrm{STr}\left(e^{i (2\pi \alpha')k_z \Phi}F_{ij}D_{k}\Phi\right)=T_{\mathrm{D3}}(2\pi \alpha')^2\mathrm{STr}\left(e^{i (2\pi \alpha')k_z \Phi}D^{k}\Phi D_{k}\Phi\right)\,.
\end{equation}
It can be decomposed in three different contributions, each of which we will evaluate separately:
\begin{equation} \label{j_D1z_three_terms}
     j_{\mathrm{D1}}^{0z}(k_z) = T_{\mathrm{D3}}(2\pi \alpha')^2 \mathrm{STr}\left(e^{i (2\pi \alpha')k_z \Phi}\left(\partial^{i}\Phi \partial_{i}\Phi+ 2\partial^{i}\Phi [A_i,\Phi]+ [A_i,\Phi] [A^i,\Phi] \right)\right)\,.
\end{equation}

The first component of the symmetrised trace, $e^{i (2\pi \alpha')k_z \Phi}\partial^{k}\Phi \partial_{k}\Phi$, only involves the scalar field (which is in the Cartan sub-algebra), and so is a diagonal matrix. The expression is straightforward to evaluate: 
\begin{align}
     \mathrm{STr}\left(e^{i (2\pi \alpha')k_z \Phi}\left(\partial^{i}\Phi \partial_{i}\Phi\right)\right)
     =\left(\left(\partial_i \phi_1\right)^2 e^{i (2\pi \alpha')k_z \phi_1}+\dots + \left(\partial_i \phi_N\right)^2 e^{i (2\pi \alpha')k_z \phi_N}\right)\,.
\end{align}

The second contribution of \eqref{j_D1z_three_terms} consists of a diagonal matrix, $2e^{i (2\pi \alpha')k_z \Phi}\partial^{i}\Phi$, multiplied with a matrix with vanishing entries in the diagonal, $[A_i,\Phi]$ \eqref{eq:Commutator covariant derivative}. As such, the product of the two matrices has vanishing entries in the diagonal, and the trace is zero. 

To evaluate the third and last term of \eqref{j_D1z_three_terms}, we first series-expand the exponential
\begin{align}
\label{eq:current D1 9}
    \str\left(e^{i (2\pi \alpha')k_z \Phi}[A_i,\Phi] [A^i,\Phi] \right)=\sum_{p=0}^\infty \frac{(i (2\pi \alpha')k_z)^p}{p!} \str \left(\Phi^p[A_i,\Phi] [A^i,\Phi] \right) \,.
\end{align}
Each symmetrised trace has $p+2$ elements: two of them are $[A_i,\Phi]$ and $p$ of them are $\Phi$. Thus, the symmetrised trace is comprised of $\binom{p+2}{2}=\frac{(p+1)(p+2)}{2}$ terms, each of the form $\tr \left( \Phi^r [A_i,\Phi] \Phi^{p-r-s} [A_i,\Phi] \Phi^s \right)$, with $0\leq r,s \leq p$. By cyclicity of the trace, one gets
\begin{equation}
    \str\left(\Phi^p[A_i,\Phi] [A^i,\Phi] \right) = \frac{2}{(p+1)(p+2)} \sum_{m=0}^{p} (m+1) \tr \left([A^i,\Phi] \Phi^{p-m}[A_i,\Phi] \Phi^{m}\right) \,.
\end{equation}
Then, using the expression \eqref{eq:Commutator covariant derivative} for $[A_i,\Phi]$, one has
\begin{align}
\label{eq:Trace double commutator}
    \mathrm{Tr}\left([A^i,\Phi] \Phi^{p-m}[A_i,\Phi] \Phi^{m}\right) &= 
    \sum_{a,b,c,d=1}^N (\phi_a-\phi_b)(\phi_c-\phi_d)A_i^{ab}A^{i,\,cd}\, \mathrm{Tr}\left(E_{ab} \Phi^{p-m}E_{cd} \Phi^{m}\right) \nonumber\\ 
    & = \sum_{a,b=1}^N -(\phi_a-\phi_b)^2 A_i^{ab}A^{i,\,ba}\phi_{a}^{m}\phi_b^{p-m}\,.
\end{align}

At this point, there are no more subtleties in the computation, and we can plug the expression \eqref{eq:Trace double commutator} in \eqref{eq:current D1 9} and compute the sum. We obtain:
\begin{align}
    \mathrm{STr}\left(e^{i (2\pi \alpha')k_z \Phi}[A_i,\Phi] [A^i,\Phi] \right)=~ 
    &\biggl(\frac{2}{(i2\pi \alpha'k_z)^2}\left(-e^{i2\pi \alpha'k_z\phi_a} + e^{i2\pi \alpha'k_z \phi_b}\right) \nn \\
    &\hspace{6em} + \frac{2(\phi_a-\phi_b)}{i 2\pi \alpha'k_z}e^{i2\pi \alpha'k_z \phi_a}\biggr)A_{i}^{ab}A^{i,\,ba}\,.
\end{align}
Because  $A_{i}^{ab}A^{i, \, ba}$ is symmetric with respect to $(a,b)$, while $-e^{i2\pi \alpha'k_z\phi_i}+e^{i2\pi \alpha'k_z \phi_j}$ is antisymmetric, the first term of the above equation vanishes.

Combining all results into \eqref{j_D1z_three_terms}, we obtain the total current:
\begin{equation}
    j_{\mathrm{D1}}^{0z}(k_z)=T_{\mathrm{D3}}(2\pi \alpha')^2 \left(\sum_{a}\left(\partial_p \phi_a\right)^2 e^{i (2\pi \alpha')k_z \phi_a} +\frac{2(\phi_a-\phi_b)}{i 2\pi \alpha'k_z}e^{i(2\pi \alpha')k_z\phi_a}A_{p}^{ab}A^{p,\,ba}\right)\,.
\end{equation}
\bibliographystyle{JHEP}
\bibliography{btwbranes}

\end{document}